\documentclass[%
 reprint,
 amsmath,amssymb,
 prstab,
floatfix,
]{revtex4-2}

\usepackage{graphicx}% Include figure files
\usepackage{dcolumn}% Align table columns on decimal point
\usepackage{bm}% bold math
\usepackage{cleveref}
\usepackage{url}
\usepackage{float}
\begin{document}

\preprint{APS/123-QED}

\title{Analysis of photoinjector transverse phase space in action and phase coordinates}% 

\author{Houjun Qian}
\email{houjun.qian@desy.de}
\affiliation{Deutsches Elektronen-Synchrotron, 15738 Zeuthen, Germany}%

\author{Mikhail Krasilnikov}
\affiliation{Deutsches Elektronen-Synchrotron, 15738 Zeuthen, Germany}%

\author{Zakaria Aboulbanine}
\affiliation{Deutsches Elektronen-Synchrotron, 15738 Zeuthen, Germany}%

\author{Gowri Adhikari}
\affiliation{Deutsches Elektronen-Synchrotron, 15738 Zeuthen, Germany}%

\author{Namra Aftab}
\affiliation{Deutsches Elektronen-Synchrotron, 15738 Zeuthen, Germany}%

\author{Prach Boonpornpras}
\affiliation{Deutsches Elektronen-Synchrotron, 15738 Zeuthen, Germany}%

\author{Georgi Georgiev}
\affiliation{Deutsches Elektronen-Synchrotron, 15738 Zeuthen, Germany}%

\author{James Good}
\affiliation{Deutsches Elektronen-Synchrotron, 15738 Zeuthen, Germany}%

\author{Matthias Gross}
\affiliation{Deutsches Elektronen-Synchrotron, 15738 Zeuthen, Germany}%

\author{Christian Koschitzki}
\affiliation{Deutsches Elektronen-Synchrotron, 15738 Zeuthen, Germany}%

\author{Xiangkun Li}
\affiliation{Deutsches Elektronen-Synchrotron, 15738 Zeuthen, Germany}%

\author{Osip Lishilin}
\affiliation{Deutsches Elektronen-Synchrotron, 15738 Zeuthen, Germany}%

\author{Anusorn Lueangaramwong}
\affiliation{Deutsches Elektronen-Synchrotron, 15738 Zeuthen, Germany}%

\author{Raffael Niemczyk}
\affiliation{Deutsches Elektronen-Synchrotron, 15738 Zeuthen, Germany}%

\author{Anne Oppelt}
\affiliation{Deutsches Elektronen-Synchrotron, 15738 Zeuthen, Germany}%

\author{Guan Shu}
\affiliation{Deutsches Elektronen-Synchrotron, 15738 Zeuthen, Germany}%

\author{Frank Stephan}
\affiliation{Deutsches Elektronen-Synchrotron, 15738 Zeuthen, Germany}%

\author{Grygorii Vashchenko}
\affiliation{Deutsches Elektronen-Synchrotron, 15738 Zeuthen, Germany}%

\author{Tobias Weilbach}
\affiliation{Deutsches Elektronen-Synchrotron, 15738 Zeuthen, Germany}%

\date{\today}% It is always \today, today,

\begin{abstract}
Photoinjectors are the main high brightness electron sources for X-ray free electron lasers (XFEL). Photoinjector emittance reduction is one of the key knobs for improving XFEL lasing, so precise emittance measurement is critical. It's well known that rms emittance is very sensitive to low intensity tails of particle distributions in the phase space, whose measurement depend on the signal to noise ratio (SNR) and image processing procedures. Such sensitivities make the interpretations of beam transverse brightness challenging, leading to different emittance definitions to reduce the impact of tail particles. In this paper, transverse phase space is analyzed in action and phase coordinates for both analytical models and experiments, which give a more intuitive way to calculate the beam core brightness.
\end{abstract}

%\keywords{Suggested keywords}%Use showkeys class option if keyword display desired
\maketitle

%\tableofcontents

\section{Introduction}
Photoinjectors combine high brightness photocathode, ultrafast laser and high gradient acceleration, and deliver directly bunched beam with picoseconds duration, low energy spread and low emittance \cite{rao2014engineering}. Photoinjectors are also very flexible in controlling electron bunch pattern and phase space by varying laser parameters, and have become the main high brightness electron source for many accelerator based scientific facilities, such as X-ray free electron lasers (XFEL), ultrafast electron microscopy and energy recovery linacs \cite{pellegrini2017x, weathersby2015mega, merminga2020energy}. Transverse emittance is a key parameter for characterizing the transverse beam brightness
\begin{eqnarray}\label{eq:B4D}
  B_{\rm 4D} \propto  \frac{Q}{\varepsilon_{x}\varepsilon_{y}},
\end{eqnarray}
where $Q$ is the bunch charge, $\varepsilon_{x}$ and $\varepsilon_{y}$ are the beam emittance for two decoupled transverse planes. Short wavelength free electron lasers (FEL), especially hard X-ray FELs (XFEL), are the main drivers of ultralow emittance injectors \cite{huang2007review}. In the design stage of the first hard X-ray FEL, 1 nC with a normalized emittance of 1 \textmu m is the baseline parameter for lasing at the wavelength around 1 {\AA} \cite{nuhn2002linac, decking2013european}. Nowadays, most XFEL injectors operates with 0.25 nC with emittance below 0.4 \textmu m \cite{emma2010first, prat2020compact, decking2020mhz}. Even lower beam emittance of 0.1 \textmu m at 100 pC is persued by the development of XFEL with lower linac energy, such as compact XFELs and XFELs based on the continuous wave superconducting linac \cite{rosenzweig2020ultra, d2019compactlight, raubenheimer2018lcls}. The XFEL developments drive the photoinjector transverse brightness $B_{\rm 4D}$ up by a factor 10 in the last decades.

Beam eittance is generally defined as the phase space area, and practically is calculated by the rms emittance
\begin{eqnarray}\label{eq:rms_emit}
  \varepsilon_{rms} =  \sqrt{\sigma_x^2\sigma_{x'}^2-\sigma_{xx'}^2},
\end{eqnarray}
where $\sigma_x$ is the rms beam size, $\sigma_{x'}$ is the rms beam divergence, $\sigma_{xx'}$ is the covariance between $x$ and $x'$. It is well known that the second moment is sensitive to non-exponential decay tails of a distribution, which makes rms emittnace sensitive to tail particles in transverse phase space. In a well optimized photoinjector, beam transverse phase space is dominated by a high density Gaussian core distribution and a non-Gaussian low density tail distribution. The non-Gaussian tail distribution is mainly due to residual nonlinear space charge effect and phase space mismatch in low current tails \cite{carlsten1995space}. Such a low density distribution accounts for a small fraction of the beam, but degrades the full beam rms emittance disproportionally, which misleads the interpretation of real beam brightness according to Eq.~(\ref{eq:B4D}). As the brightness requirement of electron source   for the next generation FEL gets higher, e.g. 0.1 \textmu m emittance for 100 pC beam \cite{raubenheimer2018lcls}, proper treatment of such low density tails can be critical.

For beam profile scan based emittance measurements, such as quadrupole magnet scan, multi-screen scan, slit mask scan, beam size processing from beam images is a critical step. Due to the limited signal to noise ratio (SNR), image noise subtraction and weak signal recognition can change the low intensity tails of the beam profile projection, hence the rms beam size changes \cite{2008emit_workshop}. Different image processing methods can lead to different emittance results for the same beam. Typically, 100\% emittance and 95\% emittance, corresponding to 100\% bunch charge and 95\% bunch charge, are used. The 95\% emittance is much less sensitive to tail particles by excluding the worst 5\% particles, but the actual charge fraction in the non-Gaussian tails of the phase space is not clear. Another way to reduce the sensitivity of non-Gaussian tails in emittance calculation is to fit the beam profile with Gaussian models, which is even less sensitive to measurement SNR, but the charge fraction in the measured emittance is not clear.

In the measurement of a 2.1 MeV H$^{-}$ beam at the PIP2IT beamline, action and phase coordinates were introduced for phase space analysis, which allows for measurements of distortions and tail growth due to non-linear forces \cite{richard2020measurements}. In this paper, this approach is used for analyzing the Gaussian core and non-Gaussian tails of photoinjector beam phase spaces. The paper is organized as follows. In Sec. II, the emittance and beam brightness calculations based on the action and phase coordinates are introduced by analyzing an simplified analytical model. In Sec. III, the phase space measurements and analysis of four bunch charges (1 nC to 0.1 nC) of FEL photoinjector are presented. Finally, a summary is given in Sec. IV.

\section{Beam brightness and emittance}
\label{method1}
Beam brightness scales as beam density in phase space, and is usually calculated by $Q/{\varepsilon}$ for one plane. For the same type of density distribution, e.g. a 2D Gaussian distribution, then $Q/{\varepsilon}$ is linearly proportional to both peak and average beam density in the phase space, therefore $Q/{\varepsilon}$ can be used as a figure of merit to compare brightness between different beams. In reality, photoinjector beam is so 'cold' that the transverse phase space is not fully thermalized to become a pure Gaussian distribution. The phase space density distribution type varies between different beams, so $Q/{\varepsilon}$ can be misleading for comparing beam brightness. 

\subsection{100\% emittance and reduced emittance}
\begin{figure}[!thp]
   \includegraphics*[width=0.8\columnwidth]{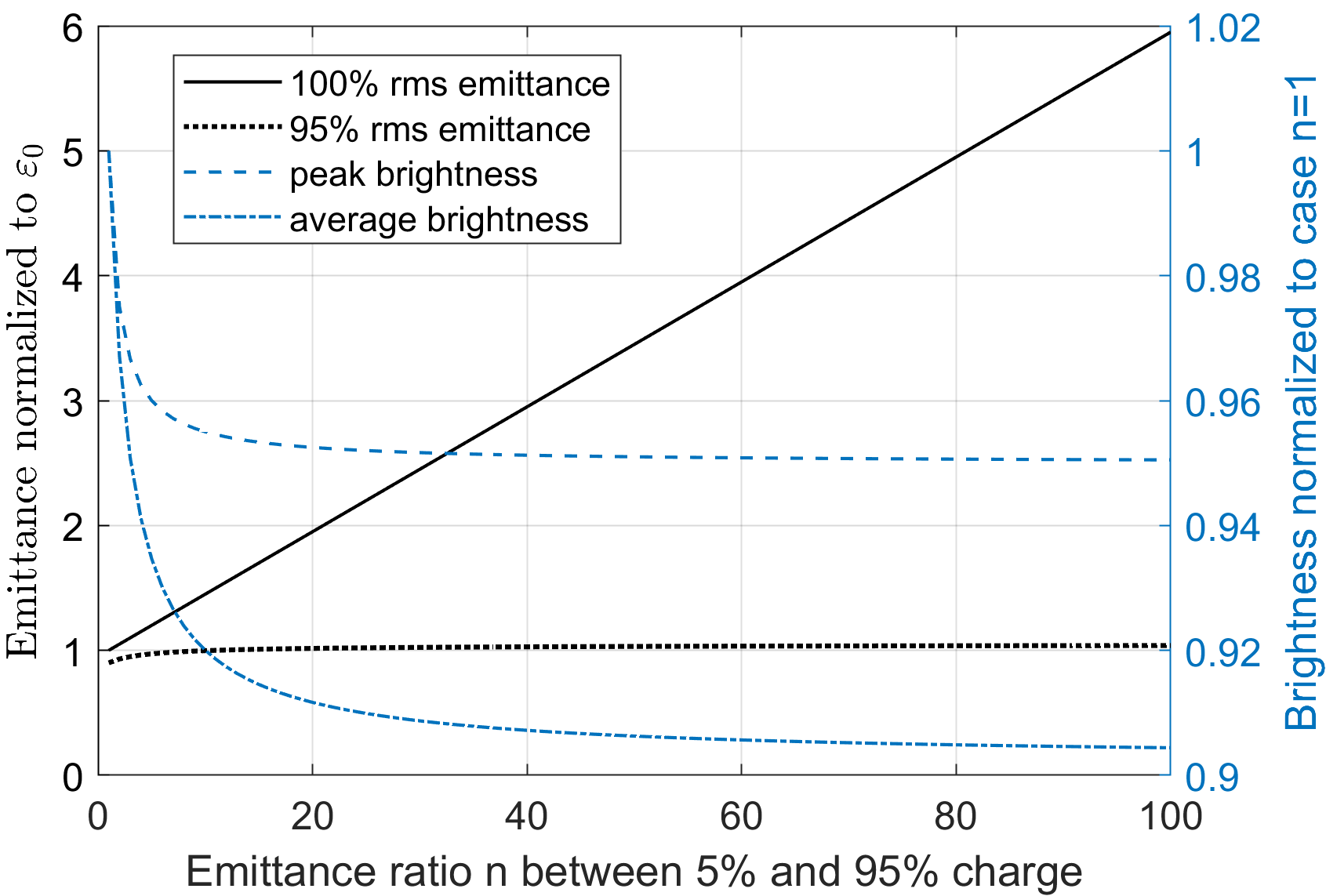}
   \caption{\label{fig:emit_ratio}RMS emittance and beam brightness vs 5\% charge emittance degradation according to Eq.~(\ref{eq:95emit}).}
\end{figure}

Assume the following phase space density distribution
\begin{eqnarray}\label{eq:95emit}
\begin{aligned}
f(x, x') =& \frac{0.95Q}{2\pi\varepsilon_0}\exp(-\frac{x^2+x'^2}{2\varepsilon_0})\\
&+\frac{0.05Q}{2\pi{n}\varepsilon_0}\exp(-\frac{x^2+x'^2}{2{n}\varepsilon_0}),
\end{aligned}
\end{eqnarray}
where 95\% charge is distributed in a Gaussian phase space with rms emittance of $\varepsilon_0$, and the other 5\% charge has a rms emittance of $n\varepsilon_0$. For simplification, the two density distributions have the same twiss parameters. The rms emittance and beam brightness of the 100\% charge is shown in Fig.~\ref{fig:emit_ratio}, in which peak and avearge beam brightness is defined as follows,
\begin{eqnarray}\label{eq:Bpeak}
B_{\rm peak} = \max{f(x, x')},
\end{eqnarray}
\begin{eqnarray}\label{eq:Bave}
B_{\rm ave} = \frac{1}{Q}\int \ f(x, x') \,dQ.
%= \frac{1}{Q}\int \ f(x, x')^2 \,dx \ dx'
\end{eqnarray}
Figure~\ref{fig:emit_ratio} shows the peak and average beam brightness reductions converge to about 5\% and 10\% respectively after the continuous degradation of the 5\% charge, but the 100\% rms emittance growth does not converge, which is misleading for understanding the real beam brightness. In comparison, the 95\% emittance is not sensitive to the 5\% charge in the tails. This example demonstrates the 100\% rms emittance is very sensitive to tail particles in the phase space, but the actual beam brightness is much less sensitive. An emittance of reduced charge can be used to reflect the real beam brightness. In this case, we know from Eq.~(\ref{eq:95emit}) that about 5\% charge is in the tails of the phase space, otherwise it's also not intuitive to find the reduced emittance with a proper charge cut to reflect the beam brightness. 

Different methods have been used to measure the reduced emittance. The most intuitive way is to measure the transverse phase space first, and then calculate the reduced emittance vs charge cut \cite{MK2012, gulliford2013demonstration}. For space charge dominated beam, the phase space can be measured by slit mask scanning transversely through the beam \cite{MK2012}. For high energy beam, the phase space is measured by tomography \cite{lohl2005measurements}, which is based on beam profile projection vs phase space rotation by beamline optics variations, such as quadrupole magnet scan or multi-screen technique. In addition, the reduced rms emittance can also be fitted with reduced rms beam size vs beamline optics. If the beam is dominated by a Gaussian distribution, then reduced beam rms size can be fitted with a Gaussian distribution to cut the tail particles \cite{prat2014emittance}. Tail particles can also be removed directly in the 2D transverse beam profile or 1D projected beam profile \cite{lohl2005measurements, 2008emit_workshop}. The problem is, tail particles in the transverse profile is not necessarily tail particles in the transverse phase space, so it's only an approximation.

\subsection{Phase space analysis in the action and phase coordinates}
Phase space can be expressed in different coordinates, which are of advantage for different analysis purposes. The following formula shows three coordinate systems for transverse phase space \cite{stupakov2018classical},
\begin{eqnarray}\label{eq:coordinates}
  \begin{pmatrix}
   x_{\scriptscriptstyle \rm N}\\    x'_{\scriptscriptstyle \rm N}
 \end{pmatrix}=
 \begin{pmatrix}
   \frac{1}{\sqrt{\beta}} & 0\\
   \frac{\alpha}{\sqrt{\beta}} & \sqrt{\beta}
 \end{pmatrix}
\begin{pmatrix}
   x\\   x'
 \end{pmatrix}=\sqrt{2J}
\begin{pmatrix}
   \cos \phi\\   -\sin \phi
 \end{pmatrix},
\end{eqnarray}
\begin{eqnarray}\label{eq:action}
  J=\frac{x_{\scriptscriptstyle {\rm N}}^2+{x'}_{\scriptscriptstyle {\rm N}}^2}{2}=\frac{\gamma x^2+2\alpha x x'+\beta {x'}^2}{2},
\end{eqnarray}
\begin{eqnarray}\label{eq:angle}
  \phi=-\arctan \frac{x'_{\scriptscriptstyle \rm N}}{x_{\scriptscriptstyle \rm N}}=-\arctan \frac{\alpha x +\beta x'}{x},
\end{eqnarray}
where $x$ and $x'$ are position and angle coordinates, $x_{\scriptscriptstyle \rm N}$ and $x'_{\scriptscriptstyle \rm N}$ are normalized coordinates, $J$ and $\phi$ are action and phase coordiantes, $\beta$, $\alpha$ and $\gamma$ are twiss parameters of the phase space. Based on Eq.~(\ref{eq:action}), rms emittance can also be calculated by
\begin{eqnarray}\label{eq:rms_emit_action}
  \varepsilon_{\rm rms}=\left \langle J \right \rangle=\frac{1}{Q}\int \ J \,dQ,
\end{eqnarray}
i.e. the rms emittance of the beam is the average of action of all particles. While $x$ and $x'$ describe the beam motion in real space, the other two coordinates are better for describing the beam motion in phase space and beam brightness. Since $J$ is proportional to phase space area, it is a better coordinate system to describe beam brightness. For example, the Gaussian phase space can be described in the following two forms:
\begin{eqnarray}\label{eq:Gauss}
\begin{aligned}
   dQ &= \frac{Q}{2\pi\varepsilon_0}\exp(-\frac{\gamma x^2+2\alpha x x'+\beta x'^2}{2\varepsilon_0})dxdx'\\
     &= \frac{Q}{\varepsilon_0}\exp(-\frac{J}{\varepsilon_0})dJ 
\end{aligned}.
\end{eqnarray}
In the action and phase coordinates, the 2D density distribution reduces to a 1D exponential distribution, which is a straight line in the log scale. This makes it much easier to distinguish the Gaussian core and the tail particles in the phase space. The peak brightness is $Q/\varepsilon_0$, and the rms emittance $\varepsilon_0$ is inversely proportional to the slope of the line in the log scale.

\begin{figure}[!thp]
   \includegraphics*[width=0.8\columnwidth]{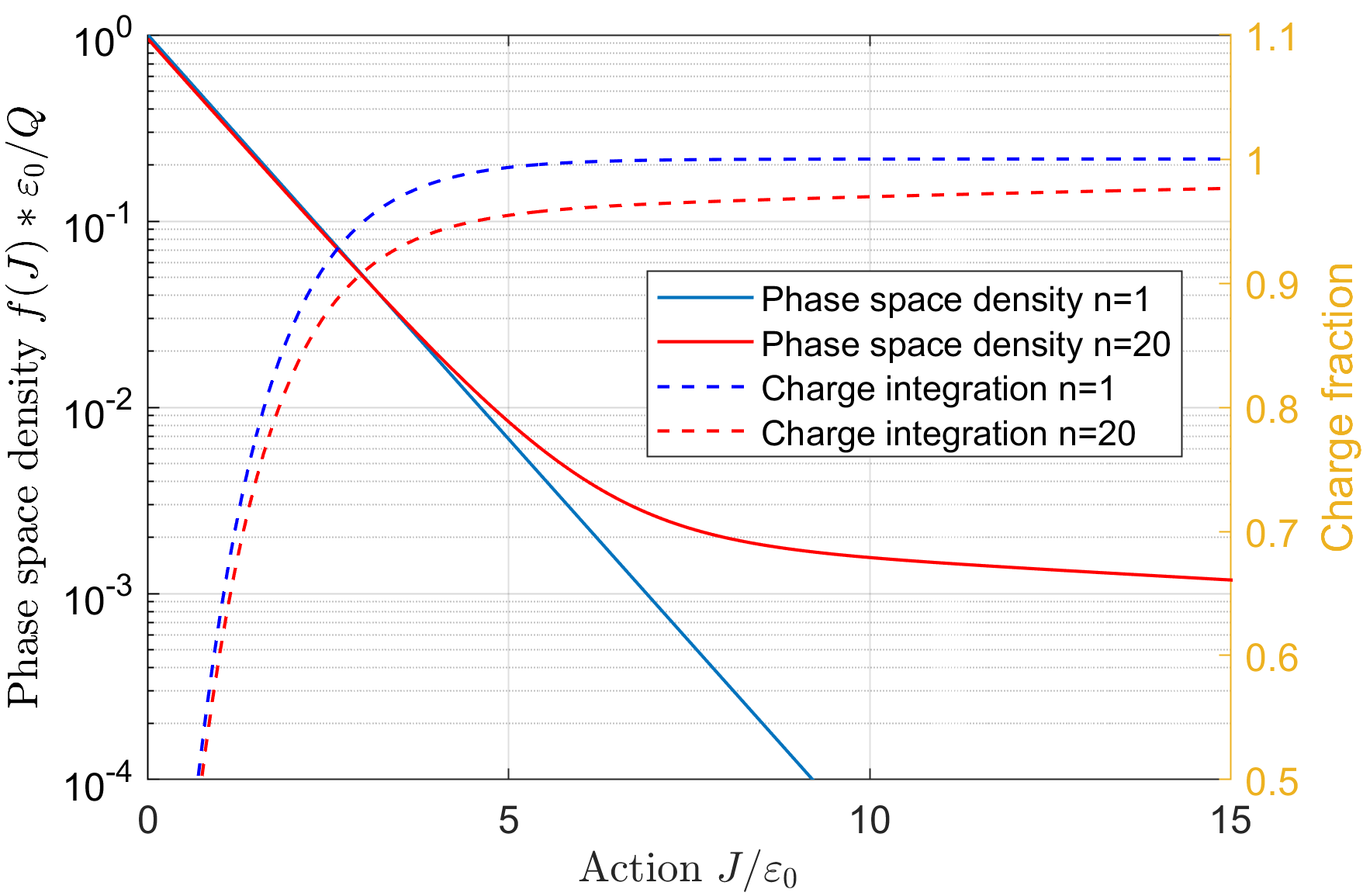}
   \caption{\label{fig:emit_ratio2}RMS emittance and beam brightness vs 5\% charge emittance degradation according to Eq.~(\ref{eq:95emit_action}).}
\end{figure}

Let us use Eq.~\ref{eq:95emit} as an example, it can be rewritten in the action and phase coordinates as,
\begin{eqnarray}\label{eq:95emit_action}
f(J) = \frac{0.95Q}{\varepsilon_0}\exp(-\frac{J}{\varepsilon_0})
+\frac{0.05Q}{n\varepsilon_0}\exp(-\frac{J}{n\varepsilon_0}).
\end{eqnarray}
The pure Gaussian mode (n=1) and the mixed mode example (n=20) are shown in Fig.~\ref{fig:emit_ratio2}. Now it's easily recognized that the mixed mode case consists of a Gaussian core and a tail. The Gaussian core has the same rms emittance as the pure Gaussian mode with a charge fraction of about 95\%. Even without prior knowledge of Eq.~(\ref{eq:95emit_action}), the charge fraction of the Gaussian core can be easily calculated in the action and phase coordinates. By fitting the Gaussian core of the mixed mode to Eq.~(\ref{eq:Gauss}), both the peak brightness and the Gaussian core emittance can be extracted, and their product gives the charge in the Gaussian core. In this paper, the Gaussian core of the phase space is fitted to Eq.~(\ref{eq:Gauss}) in the action-phase coordinates, and the fitted Gaussian mode defines the core emittance $\varepsilon_{\rm core}$ and the core charge $Q_{\rm core}$. For a pure Gaussian distribution in Eq.~(\ref{eq:Gauss}), the core emittance is the same as 100\% rms emittance $\varepsilon_0$, and the core charge is the same as full charge $Q$.

There exists another definition for core emittance and core charge in the literature \cite{bazarov2009maximum},
\begin{eqnarray}\label{eq:core_emit_cornell}
\varepsilon_{\rm core, 2} = Q\cdot \frac{d\varepsilon(Q)}{dQ}\bigg|_{Q = 0}=\varepsilon(Q_{\rm core, 2}).
\end{eqnarray}
Applying the above definition Eq.~(\ref{eq:core_emit_cornell}) to a Gaussian distribution in Eq.~(\ref{eq:Gauss}), the core emittance and core charge are $\varepsilon_0/2$ and 0.715Q respectively. This is clearly different from our definition.

\section{Measurements of 100\% emittance and core emittance}
\begin{figure}[!tbp]
   \includegraphics*[width=0.8\columnwidth]{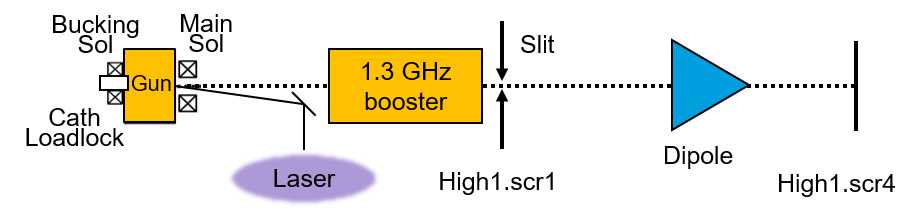}
   \caption{\label{fig:PITZ_EMSY1}Schematic plot of the projected emittance measurement beamline at PITZ, High1.scr1 and High1.scr4 are two screen stations for slit-scan emittance measurements. Slit is installed at High1.scr1 and beamlet images are measured at High1.scr4. Dipole magnet is for beam energy measurement.}
\end{figure}

PITZ is a photoinjector R\&D facility for developping high brightness electron source for pulsed superconducting linac based Free Electron Laser in Hamburg (FLASH) and European XFEL \cite{FS2010, MK2012}. The three facilities use the same type 1.3 GHz RF gun \cite{BD1997, paramonov2017design}, featuring both high cathode gradient (60 MV/m) and long RF pulse length (up to 1 ms) at 10 Hz repetition rate. The schematic plot of the projected emittance measurement beamline at PITZ is displayed in Fig.~\ref{fig:PITZ_EMSY1}. Following the gun, a normal conducting booster linac accelerates the beam to about 20 MeV. The projected emittance diagnostic section includes a dipole magnet for beam momentum measurement and a slit-scan system for emittance measurement since the beam is still space charge dominated. The photoelectrons are generated by 257.5~nm UV photoemission from $\rm{Cs_2Te}$ cathode. The typical quantum efficiency and intrinsic emittance of the fresh $\rm{Cs_2Te}$ cathode are 10-30\% and 1 \textmu m/mm respectively \cite{lederer2018cs2te, huang2020single, huang2019test}. The UV laser can be shaped both spatially and temporally for emittance optimization \cite{will2008generation, gross2019emittance}. Several European XFEL injector working points were studied at PITZ for both temporally flattop and Gaussian laser.

\subsection{Charge cut in slit-scan due to noise floor}
The slit mask at PITZ is made of 1 mm thick tungsten mask with two opening options, 10 \textmu m and 50 \textmu m, and a motor drives the slit mask through the beam transversely \cite{staykov2005design}. The masked beam will be scattered, forming an uniform background, and a small beamlet will go through the slit opening to map the local beam divergence to a YAG screen at high1.scr4. The motor speed can be adjusted, typically between 0.1 mm/s and 0.5 mm/s, and the high1.scr4 camera is triggered at 10 Hz, so the beam is sampled with a step size between 10 \textmu m and 50 \textmu m. Both the beamlet signal and the background signal are recorded during slit-scan with laser shutter open and closed, respectively. The data acquisition process is controlled by a software named 'FASTSCAN' and finishes within 1 minute for one emittance measurement.

The slit-scan based emittance measurements are limited by three factors: space charge effect, spatial-angular resolution, and signal to noise ratio \cite{niemczyk2021subpicosecond}. The 50 \textmu m slit width will allow higher beamlet signal with reduced number of bunches in one RF pulse, but the space charge effect will overestimate the emittance by about 10\%. The 10~\textmu m slit width is used in this paper, and the space charge effect is negligible in emittance reconstruction. With the 10~\textmu m slit at High1.scr1 and the 50~\textmu m point spread spatial resolution for YAG screen at High1.scr4, the drift distance in between is about 3 meter, so the minimum rms emittance that can be resolved by slit-scan at 20 MeV is about 7 nm, which is more than enough for resolving typical XFEL injector emittance. In the end, the signal to noise ratio dominates the emittance measurement results. Before the slit-scan data acquisition, the number of bunches in one RF pulse is always increased until the maximum beamlet image on high1.scr4 gets close to saturation for a 14 bit camera, and then bunch train length is kept during the slit-scan. When slit moves to the transverse edges of the beam, the beamlet signal gets very weak and even below the noise floor of the camera. Therefore, part of the beamlet signal cannot be recognized during the image data processing, and the final reconstructed phase space does not include the full bunch charge, i.e. the rms emittance is not the 100\% emittance. One indicator is that the rms beam size measured at the slit location by a YAG screen is bigger than rms beam size projection from the reconstructed beam phase space \cite{rimjaem2010generating}.
 
\begin{figure}[!thp]
   \includegraphics*[width=0.8\columnwidth]{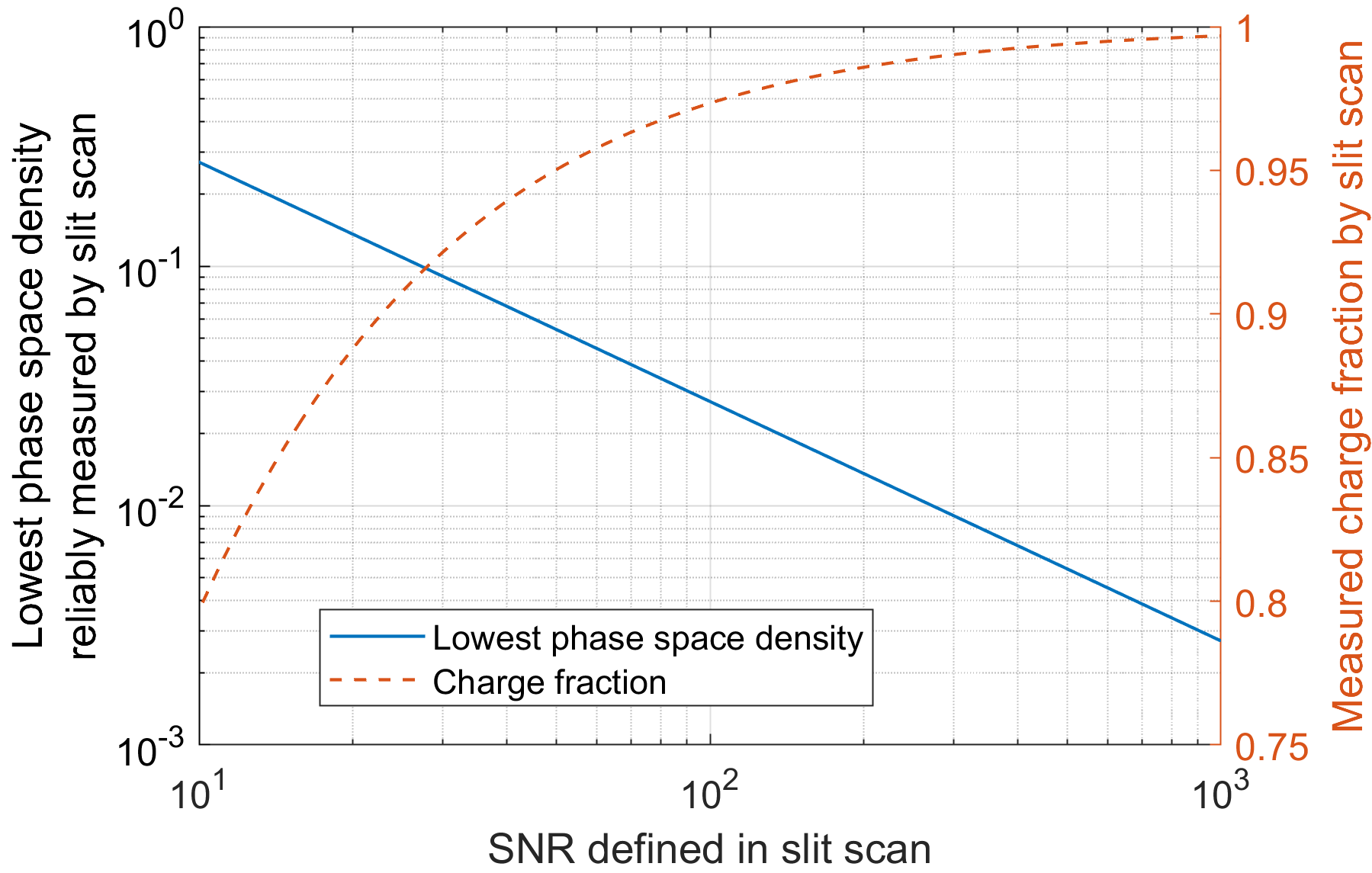}
   \caption{\label{fig:Q_fraction}Threshold phase space density and charge fraction of slit-scan vs SNR.}
\end{figure}

The Gaussian beam in Eq.~(\ref{eq:Gauss}) is used as an example for the slit-scan analysis. Assume there is no coupling between x and y plane, after the slit cut, the beamlet distribution at the measurement screen high1.scr4 is determined by the beam angle distribution $x'$ at High1.scr1 and beam spatial distribution $y_2$ at High1.scr4,
 %
%\begin{eqnarray}\label{eq:beamlet1}
%\frac{\partial Q}{\partial x'}= \frac{Q\Delta_s}{2\pi\varepsilon_0}e^{-\frac{\gamma %x_{s}^2+2\alpha x_s x'+\beta x'^2}{2\varepsilon_0}}
%\end{eqnarray}
%
\begin{eqnarray}\label{eq:beamlet2}
\frac{\partial^2 Q}{\partial x' \partial y_2}= \frac{Q\Delta_{\rm s}}{(2\pi)^{1.5}\varepsilon_0\sigma_{y_2}}e^{-\frac{\gamma x_{\rm s}^2+2\alpha x_{\rm s} x'+\beta x'^2}{2\varepsilon_0}-\frac{y_2^2}{2\sigma_{y_2}}},
\end{eqnarray}
where $x_{\rm s}$ and $\Delta_{\rm s}$ are the slit location and width respectively, $y_2$ and $\sigma_{y_2}$ are the vertical coordinate and rms beam size at High1.scr4 respectively. Eq.~(\ref{eq:beamlet2}) is integrated over $y_2$ to get the $x'$ distribution at $x_s$. In theory, the integration range for $y_2$ is infinite. In measurements, this is limited by the camera noise floor. When beamlet image pixel intensity is below the camera noise floor, it is recognized as zero and can't be integrated anymore. The camera noise floor in our measurements is defined as the rms camera noise ($\sigma_{\rm noise}$). Therefore the integration range of $y_2$ is constrained by
\begin{eqnarray}\label{eq:y2_range}
\frac{Q\Delta_{\rm s}}{(2\pi)^{1.5}\varepsilon_0\sigma_{y_2}}e^{-\frac{\gamma x_{\rm s}^2+2\alpha x_s x'+\beta x'^2}{2\varepsilon_0}-\frac{y_2^2}{2\sigma_{y_2}}}>\sigma_{\rm noise}.
\end{eqnarray}
After $y_2$ integration under the constraint in Eq.~(\ref{eq:y2_range}), the measured phase space density is
\begin{eqnarray}\label{eq:rho_measured}
\rho_{\rm measured}(x,x')=\frac{Q}{2\pi\varepsilon_0}e^{-\frac{\gamma x^2+2\alpha x x'+\beta x'^2}{2\varepsilon_0}}\rm{erf(b)}.
\end{eqnarray}
The signal to noise ratio (SNR) is defined as the ratio between maximum pixel intensity during the slit-scan and the camera rms noise. $\rm erf$ is error function, and b is defined as
\begin{eqnarray}\label{eq:b}
{\rm b}=\sqrt{\ln({\rm SNR})-\frac{\gamma x^2+2\alpha x x'+\beta x'^2}{2\varepsilon_0}}.
\end{eqnarray}
When b$>$1, $\rm erf(b)>0.84$, meaning the measured phase space density at $(x,x')$ is more than 84\% of the true density. If b$>$1 is defined as the threshold for the reliable phase space density measurement, then the threshold phase space density normalized by the peak phase space density is $2.7/\rm SNR$ based on Eq.~(\ref{eq:b}). If Eq.~(\ref{eq:beamlet2}) is integrated over $x_s$, $x'$ and $y_2$ within the constraint of Eq.~(\ref{eq:y2_range}), then the charge in the measured phase space is
\begin{eqnarray}\label{eq:Q_measured}
Q_{\rm measured}=(\rm erf(\sqrt{\ln \rm SNR})-\frac{2\sqrt{\ln \rm SNR}}{\rm SNR \sqrt{\pi}})Q.
\end{eqnarray}
The lowest reliable phase space density and charge fraction in the measured phase space versus the SNR are displayed in Fig.~\ref{fig:Q_fraction}.
When the phase space is not Gaussian and has more particles in the tails, e.g. like Eq.~(\ref{eq:95emit}), then the charge fraction measured by the slit-scan is even lower than Eq.~(\ref{eq:Q_measured}). Therefore the measured rms emittance by slit-scan is not 100\% rms emittance, and it is sensitive to the SNR.

\subsection{1 nC beam with temporal flattop laser}
\begin{figure}[!tbp]
   \includegraphics*[width=0.8\columnwidth]{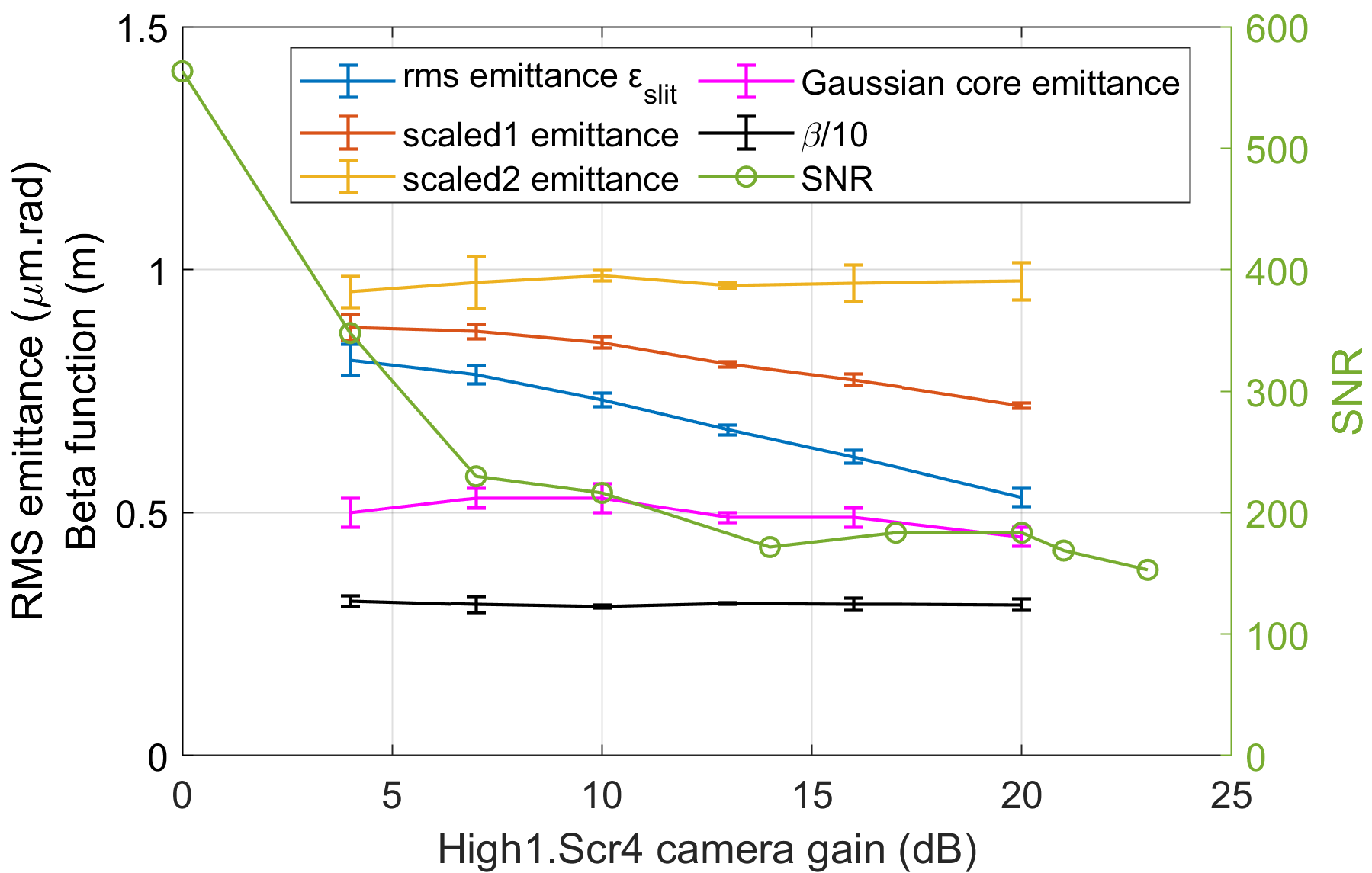}
   \caption{\label{fig:emit_vs_gain2}Slit-scan vs camera gain.}
\end{figure}

\begin{figure}[!tbp]
   \includegraphics*[width=0.8\columnwidth]{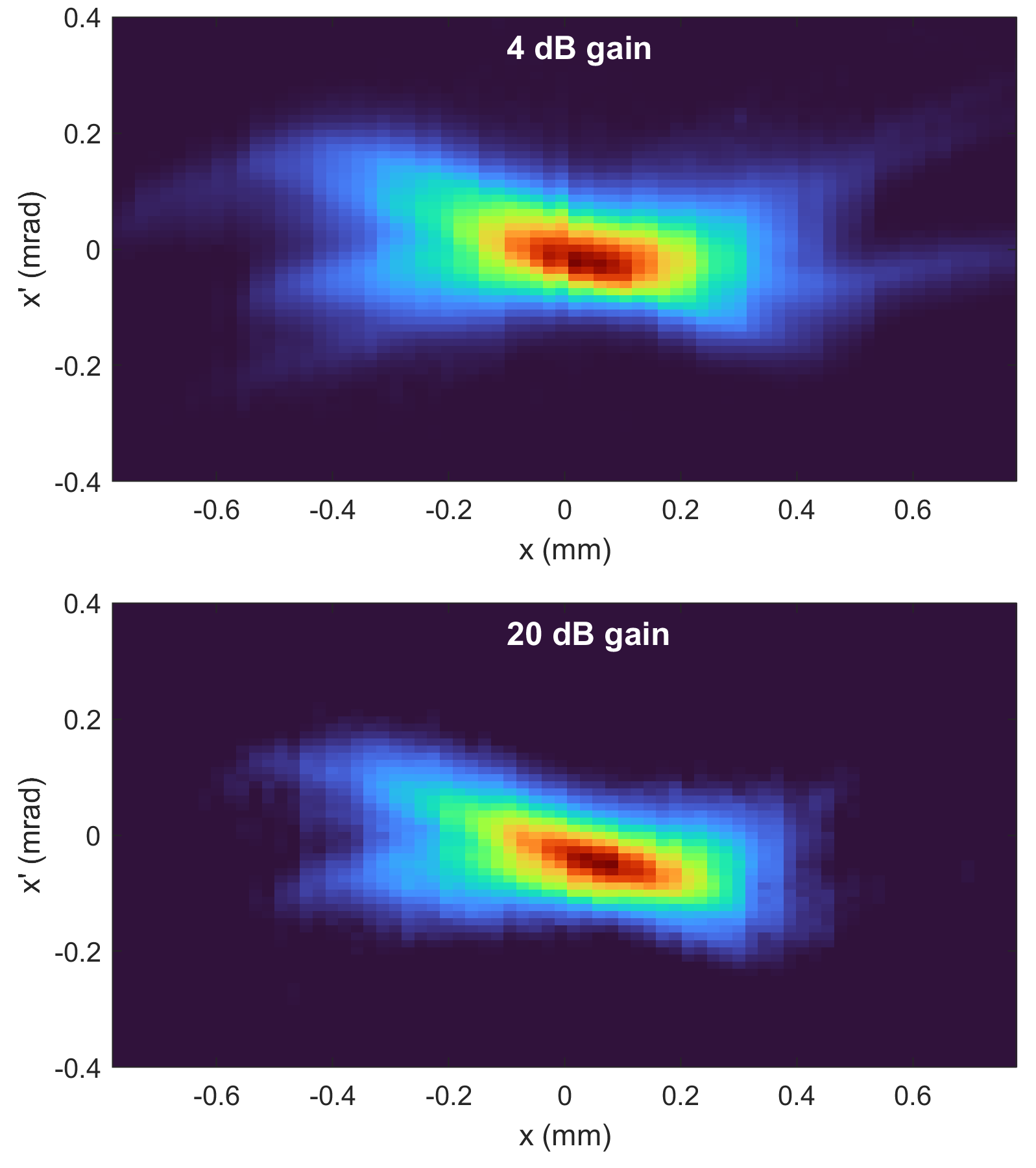}
   \caption{\label{fig:4_20dB}1 nC beam measured with two camera gains, unscaled rms emittance are 0.81 \textmu m at 4 dB and 0.53 \textmu m at 20 dB, respectively.}
\end{figure}

The 1 nC beam rms emittance measurement is used as an example to show the SNR dependence. The emittance optimization starts with beam based alignment between laser, gun, solenoid and booster \cite{krasilnikov2005beam}. The UV laser is shaped for emittance optimization. For this measurement, laser is temporally flattop with 17 ps FHWM, and is spatially flattop with an optimum diameter of 1.3 mm. Emittance is measured versus both solenoid scan and laser diameter scan. The SNR is varied by High1.scr4 camera gain, and the maximum beamlet pixel intensity during slit-scan is kept close to the camera saturation level (4000) by varying the bunch numbers in one RF pulse. The 1 nC beam emittance vs camera gain is shown in Fig.~\ref{fig:emit_vs_gain2}. When camera gain is reduced from 20 dB to 4 dB, camera rms noise reduces, and SNR increases. With higher SNR, slit-scan measures more tail particles in the phase space, as shown in Fig.~\ref{fig:4_20dB}, therefore the measured rms emittance increases from 0.53 \textmu m at 20 dB to 0.81 \textmu m at 4 dB.

In order to recover the 100\% emittance, the measured emittance is scaled up based on the rms beam size at High1.scr1. The scaling factor is defined as \cite{rimjaem2010generating}
\begin{eqnarray}\label{eq:scaling}
s = \frac{\sigma_{\scriptscriptstyle \rm YAG}}{\sigma_{\rm slit}},
\end{eqnarray}
where $\sigma_{\scriptscriptstyle \rm YAG}$ is rms beam size measured by a YAG screen at High1.scr1, and $\sigma_{\rm slit}$ is the rms beam size projection from the phase space measured by slit-scan. The 100\% emittance is scaled as
\begin{eqnarray}\label{eq:scale}
\frac{\varepsilon_{100}}{\varepsilon_{\rm slit}}\varepsilon_{\rm slit}
\approx\frac{\sigma_{\scriptscriptstyle \rm YAG}^2/\beta}{\sigma_{\rm slit}^2/\beta_{\rm slit}}\varepsilon_{\rm slit}
=\frac{\beta_{\rm slit}}{\beta}s^2\varepsilon_{\rm slit},
\end{eqnarray}
where $\beta$ and $\beta_{\rm slit}$ are true and measured beam beta function at High1.scr1, $\varepsilon_{\rm slit}$ is the rms emittance measured by slit-scan. As shown in Fig.~\ref{fig:emit_vs_gain2}, the measured beam beta function is not sensitive to SNR for the flattop laser case, so $\beta_{\rm slit}/\beta$ can be approximated to 1.~Then the scaled emittance is defined as
\begin{eqnarray}\label{eq:scaled2}
\varepsilon_{\rm scaled2}=s^2\varepsilon_{\rm slit}.
\end{eqnarray}
For Gaussian laser cases, non-Gaussian tails in phase space are much larger compared to flattop laser case due to the low charge density in beam temporal tails, which are mismatched from the gun solenoid focusing and form a distribution with different twiss parameters from the temporal core particles. When these temporal tail particles have a larger beta function compared to the core particles, $\beta_{\rm slit}/\beta$ can be smaller than 1, so another scaled emittance is defined as
\begin{eqnarray}\label{eq:scaled1}
\varepsilon_{\rm scaled1}=s\varepsilon_{\rm slit}.
\end{eqnarray}
These two scaled emittance definitions are only approximations to recover 100\% emittance, and the uncertainties reduce when scaling factor is close to 1, so the SNR of the slit-scan should be enhanced to reduce the scaling factor for a more reliable 100\% emittance scaling. Figure~\ref{fig:emit_vs_gain2} shows the scaled1 and scaled2 emittance for 1 nC beam vs camera gain. As the SNR increases, the scaling factor is reduced from 1.35 at 20 dB to 1.08 at 4 dB. The scaled2 emittances are almost independent of SNR, and the average value is 0.97 \textmu m, which indicates a good scaling for the 1 nC beam 100\% emittance. The concept of scaled emittance to recover 100\% emittance also works for other emittance measurement like quadrupole magnet scans.

\begin{figure}[!thp]
   \includegraphics*[width=0.8\columnwidth]{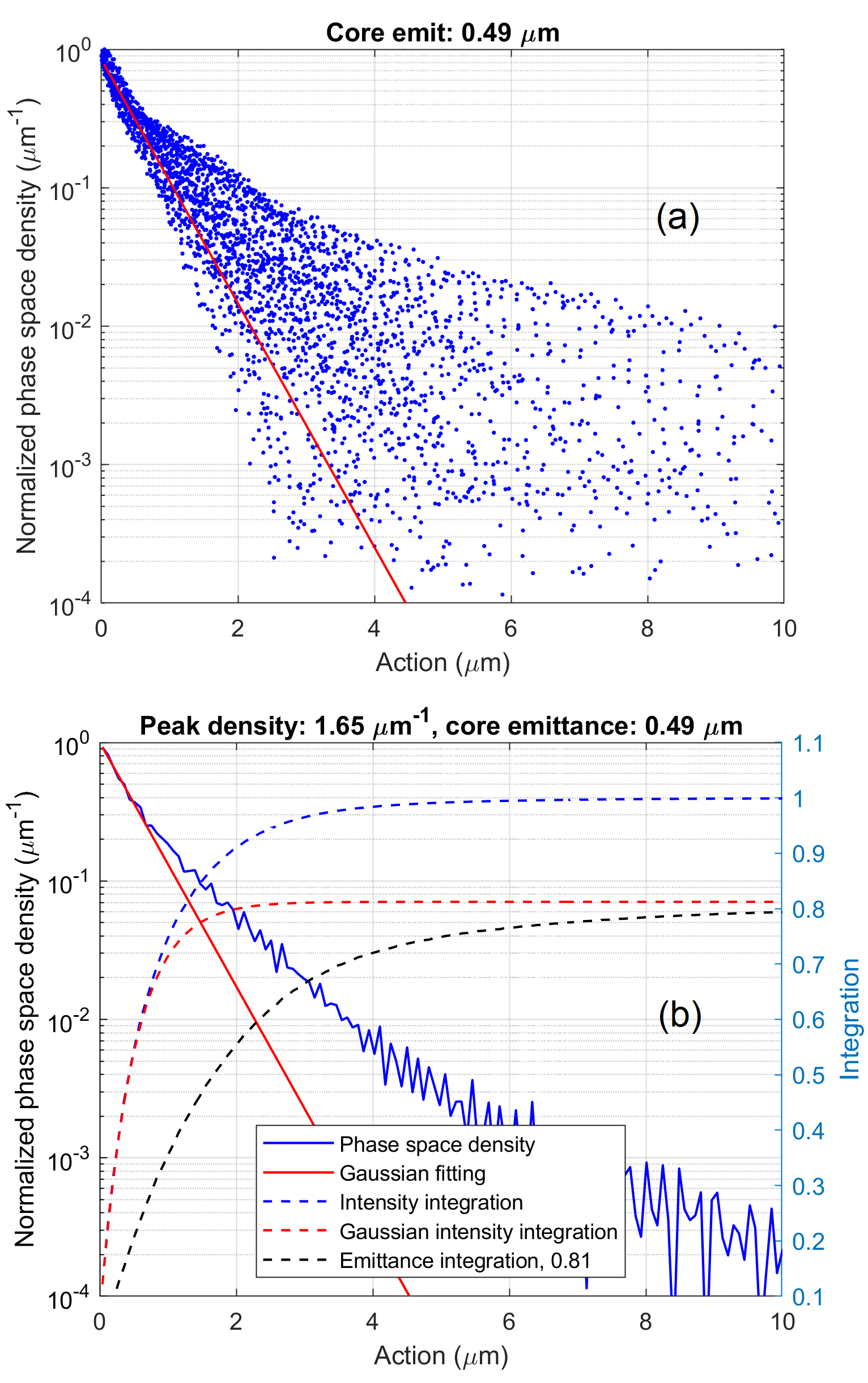}
   \caption{\label{fig:1nC_4dB_action4}Gaussian core analysis of 1 nC phase space measured at 4 dB camera gain, (a) scattering plot, (b) binning plot.}
\end{figure}

Besides scaled emittance, the Gaussian core emittance is also analyzed. First, the twiss parameters of the phase space are calculated. In order to fit better the core part of the phase space, only beam intensity above 10\% of the peak intensity of the phase space are used for twiss parameter calculations. Then, x and x' coordinates are transformed into action and phase coordinates based on Eq.~(\ref{eq:action}) and Eq.~(\ref{eq:angle}). Finally, the phase space intensities are projected to the action axis, and the Gaussian phase space core can be fitted. The 1 nC beam phase space measured with the 4 dB camera gain is transformed to action-phase coordinates,
and phase space density versus action is shown in Fig.~\ref{fig:1nC_4dB_action4}. According to Eq.~(\ref{eq:Gauss}), Gaussian phase space density is only dependent on action, not on phase, this can be useful to check the Gaussian core distribution. In Fig.~\ref{fig:1nC_4dB_action4} (a), it shows a clear correlation between density and action in the range between the peak phase space density and 30\% of the peak density, which is a Gaussian distribution. Below that range, phase space density for a defined action scatters in a large range, indicating also a phase dependence.~The Gaussian core emittance is fitted between the peak density and 30\% of the peak density, which leads to an core emittance of 0.49~\textmu m. In order to improve the statistics and to reduce the core emittance fitting dependence on twiss parameters, the scattered data points in Fig.~\ref{fig:1nC_4dB_action4} (a) are binned with an action step size of 0.08 \textmu m in Fig.~\ref{fig:1nC_4dB_action4} (b). Similar to Fig.~\ref{fig:1nC_4dB_action4} (b), a Gaussian fit between peak density and 30\% of the peak density gives a core emittance of 0.49 \textmu m and peak brightness of 1.65 $\mu\rm{m}^{-1}$, respectively. Here, the peak brightness is normalized to the total charge, 
\begin{eqnarray}\label{eq:Bpeak2}
B_{\rm peak,n}=\frac{dQ/Q}{dJ}\bigg|_{J = 0}.
\end{eqnarray}
The product of core emittance and normalized peak brightness gives the charge fraction of 81\% in the Gaussian core distribution, i.e. the solid red curve in Fig.~\ref{fig:1nC_4dB_action4} (b). The above calculations take the measured phase space charge as 100\% charge, but the SNR for the 4 dB case is about 300, so the phase space density down to the 10$^{-2}$ level is reliably measured, as shown in Fig.~\ref{fig:Q_fraction}. Based on ASTRA simulations \cite{astracode}, this corresponds to a total charge fraction of about 96\% in the measured phase space. Therefore, the total charge fraction of the core emittance of 0.49 \textmu m corresponds to 77\%. In the measured phase space, the last 4\% particles increases the emittance by 25\% (the black dash curve in Fig.~\ref{fig:1nC_4dB_action4}~(b)), which shows again the 100\% emittance is not the right parameter for beam brightness calculation. The 1 nC beam Gaussian core emittance vs slit-scan camera gain, i.e. SNR, are shown in Fig.~\ref{fig:emit_vs_gain2}. In contrast to non-scaled rms emittance, the Gaussian core emittance is almost independent of SNR. This is because the core part of the phase space has a much higher signal than the tail part of the phase space, so it is much less affected by the noise floor of the beam profile measurements. 

\subsection{Sub 1 nC beam brightness with temporal Gaussian laser}
The bunch charge of 1 nC with a flattop laser used to be the design working point for X-ray free electron lasers. Although the design goal of sub 1 \textmu m emittance was achieved for 1 nC beam in the injector, all injectors for X-ray FEL machines operate with temporal Gaussian laser and sub 1 nC charge after the FEL optimizations. The European XFEL injector is no exception to that. The standard operation charge is reduced to 250 pC with a 7 ps (FWHM) temporal Gaussian laser \cite{chen2020beam}, and both 100 pC and 500 pC were also considered \cite{Xfel250pC2020, chen2021perspectives}. The transverse phase space of the three bunch charges were also studied at PITZ with the European XFEL gun (58 MV/m) and cathode laser configurations. The emittance optimization process is similar to that described for 1 nC case. The transverse phase was measured by slit-scan with a SNR of about 300. 
\begin{table}[!htp]
\caption{\label{tab:emit_summary}Summary of different bunch charge studies at PITZ. Emittance results are geometric mean of x and y planes, $\varepsilon_{\rm slit}$ is measured by slit-scan without scaling, $\varepsilon_{\rm cath}$ is the cathode intrinsic emittance, $\varepsilon_{\rm core}$ and $Q_{\rm core}$ are the emittance and charge of the Gaussian core phase space fitted by Eq.~(\ref{eq:Gauss}), $B_{\rm 4D, peak}=(\frac{Q_{\rm core}}{\varepsilon_{\rm core}})^2/Q$.}
\begin{ruledtabular}
\begin{tabular}{cccccc}
           Q  & 1 & 0.5 & 0.25 & 0.1 & nC \\ \hline
Cathode field & 60 & \multicolumn{3}{c}{58} & MV/m  \\
Temporal shape & Flattop & \multicolumn{3}{c}{Gaussian} & /  \\
Laser duration  & 17 & \multicolumn{3}{c}{7} & ps \\
Laser diameter  & 1.3 & 1.3 & 1.0 & 0.6 & mm \\
       $I_{\rm peak}$  & 40 & 32 & 20 & 9 & A \\
$\varepsilon_{\rm cath}$    & 0.33 & 0.33 & 0.25 & 0.15 & \textmu m \\
$\varepsilon_{\rm slit}$    & 0.81 & 0.64 & 0.45 & 0.26 & \textmu m \\
$\varepsilon_{\rm core}$  & 0.49 & 0.52 & 0.37 & 0.20 & \textmu m \\
$\varepsilon_{\rm cath}/\varepsilon_{\rm core }$   & 67   & 63   & 68   & 75 & \% \\
$Q_{\rm core}/Q$   & 77   & 87   & 89   & 91 & \% \\
$Q/\varepsilon_{\rm slit}^2$  & 1.5 & 1.2 & 1.2 & 1.5 & nC/\textmu m$^2$ \\
$B_{\rm 4D, peak}$            & 2.4 & 1.4 & 1.5 & 2.0 & nC/\textmu m$^2$ \\
\end{tabular}
\end{ruledtabular}
\end{table}
\begin{figure}[!thp]
   \includegraphics*[width=\columnwidth]{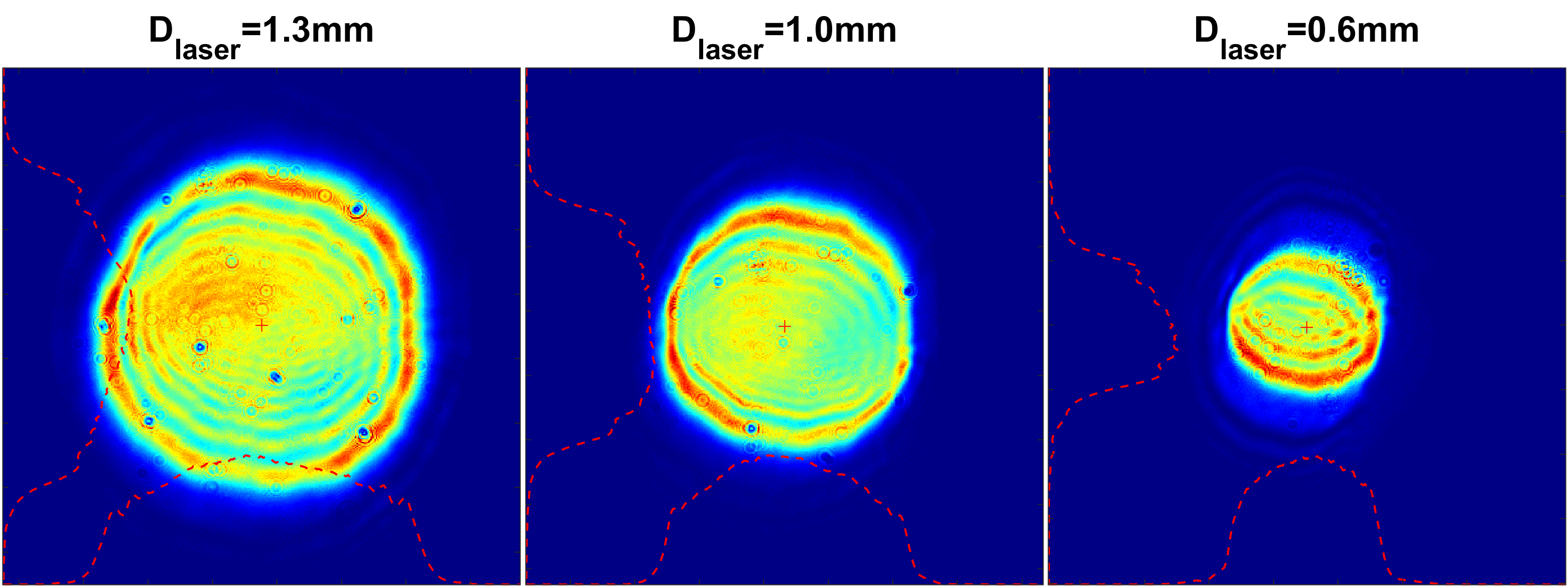}
   \caption{\label{fig:BSA}Virtual cathode laser profiles with three diameters.}
\end{figure}

\begin{figure}[!thp]
   \includegraphics*[width=0.8\columnwidth]{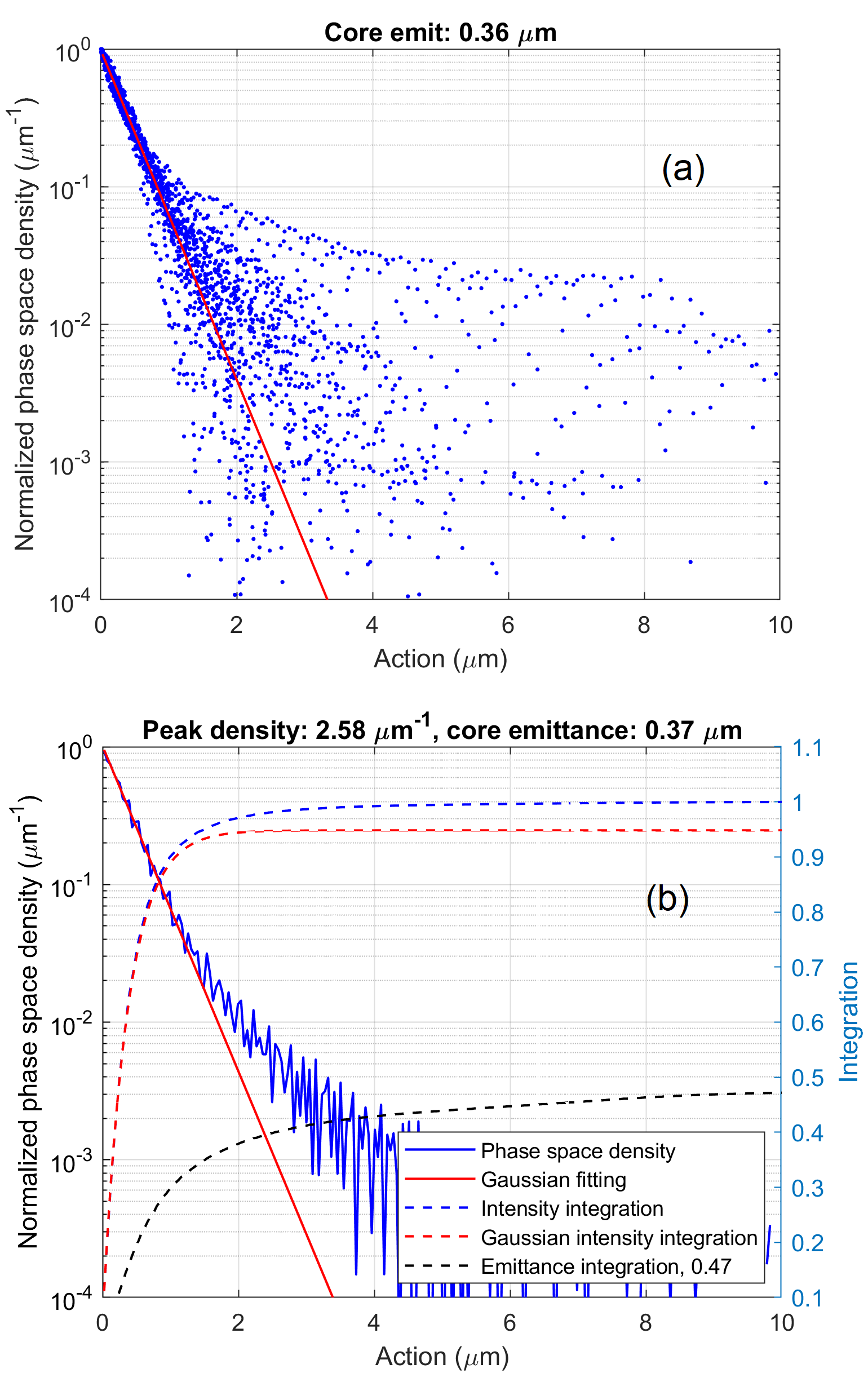}
   \caption{\label{fig:250pC_4dB_action}Gaussian core analysis of 250 pC beam phase space, (a) scattering plot, (b) binning plot.}
\end{figure}

The four bunch charge cases are summarized in Table \ref{tab:emit_summary}, and corresponding transverse laser profiles are displayed in Fig.~\ref{fig:BSA}. The phase phase projection to the action axis for 250 pC case is shown in Fig.~\ref{fig:250pC_4dB_action}. From 500 pC to 100 pC, the 4D brightness of the Gaussian core increases by 40\%, in contrast to a brightness increase of only about 21\% based on the nominal charge and nominal emittance $\varepsilon_{\rm slit}$. The 4D brightness of the 250 pC beam with Gaussian laser is 50\% higher than the original injector brightness goal of 1 nC/\textmu m$^2$, but is still about 37\% lower than the measured brightness of 1 nC beam with a flattop laser.

For all cases in Table \ref{tab:emit_summary}, the final Gaussian core emittance is dominated by the cathode thermal emittance $\varepsilon_{\rm cath}$. Without reducing the beam peak current in the injector, the beam brightness can be further enhanced by improving the cathode thermal emittance. Our Cs$_2$Te cathode typically starts with a high fresh QE above 10\%, and its life time above 1\% QE was demonstrated to be more than years in the gun \cite{lederer2018cs2te}, but its typical thermal emittance is around 1 \textmu m/mm \cite{huang2020single}, roughly a factor of 2 higher than that at SwissFEL, whose QE is roughly around 1\% \cite{prat2014thermal}. Assuming a 1\% level QE, our laser pulse energy still supports 250 pC and 500 pC operation, but the total operation life time of the cathode may become tight compared to a cathode of 10\% level QE, leading to more frequent cathode exchanges. Besides the UV cathode, Alkali-antimonide cathodes are also under R\&D at PITZ, which emits in green with a lower thermal emittance but higher dark current than Cs$_2$Te cathode \cite{Qian:471884}. Both options are under considerations for further improving the injector brightness to support lasing at even higher photon energies at European XFEL \cite{chen2021perspectives}.

\begin{figure}[!thp]
   \includegraphics*[width=0.8\columnwidth]{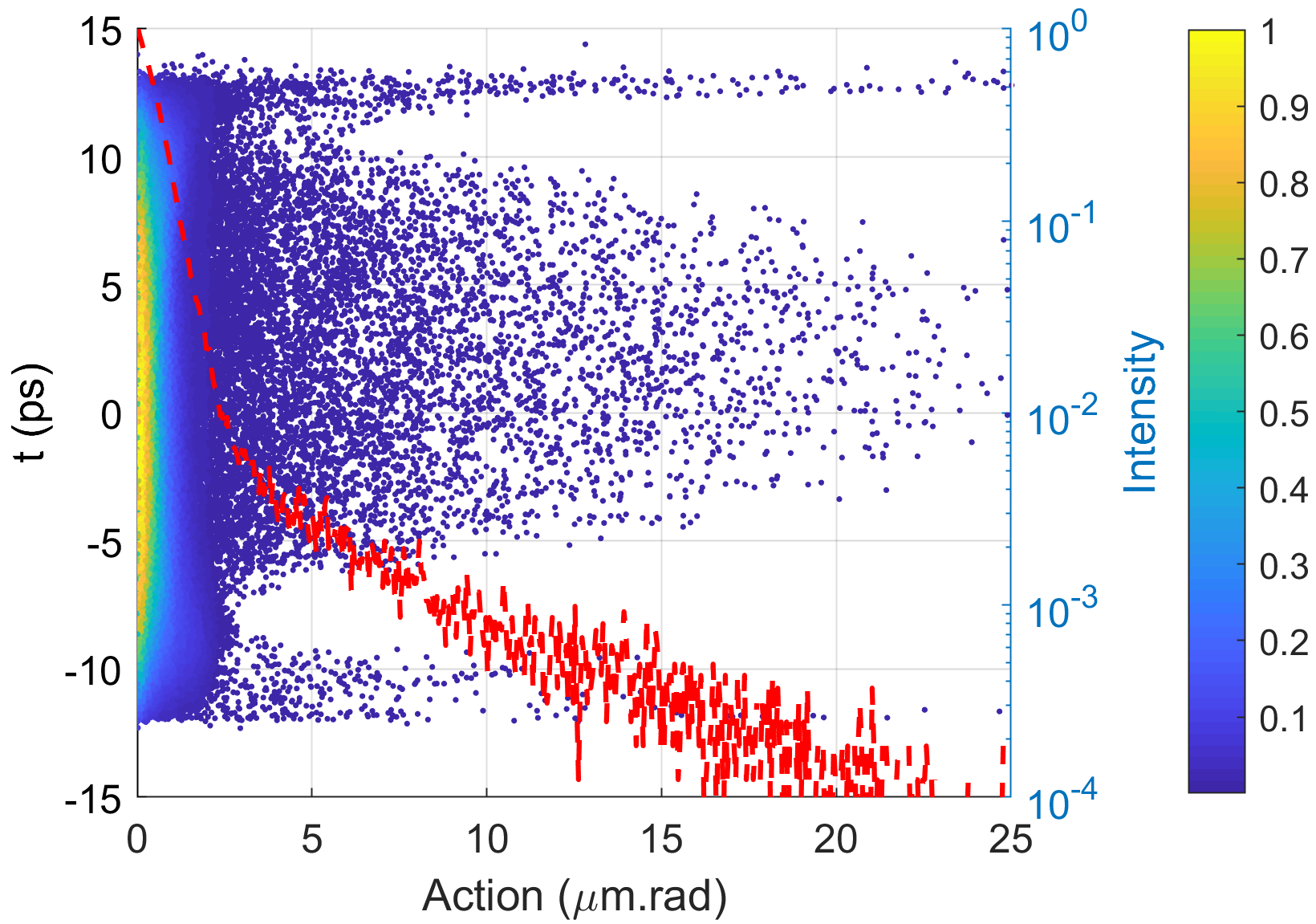}
   \caption{\label{fig:tail_1nC} Time vs action for 1 nC beam with a 17 ps flattop laser from ASTRA simulations.}
\end{figure}

\begin{figure}[!thp]
   \includegraphics*[width=0.8\columnwidth]{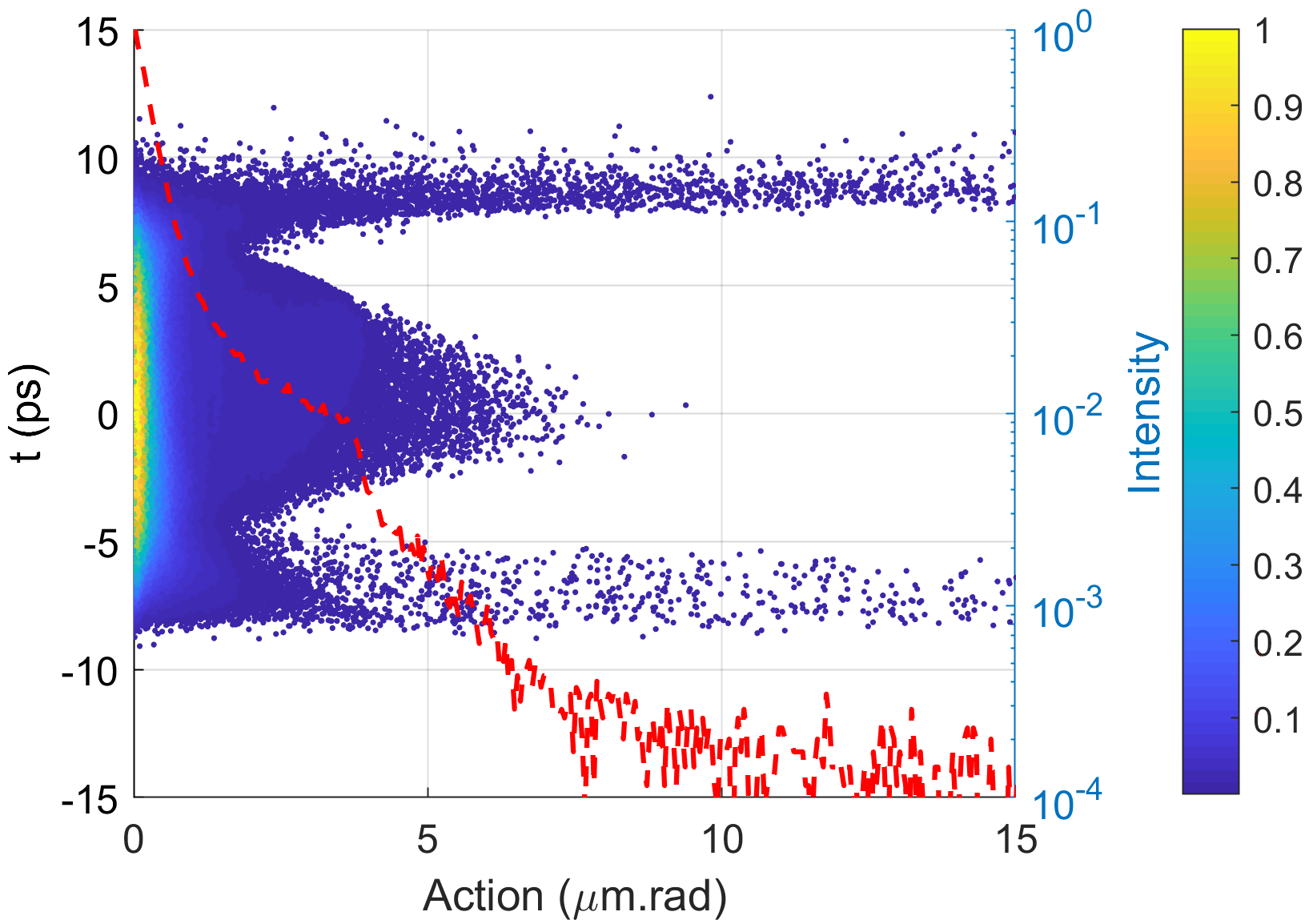}
   \caption{\label{fig:tail_250pC}Time vs action for 250 pC beam with a 7 ps Gaussian laser from ASTRA simulations.}
\end{figure}

The temporal location of the large action particles are also interesting. To measure the temporal distribution, time resolved phase space measurements are necessary, which is not included in this study. Here ASTRA simulations with ideal flattop and Gaussian laser distributions are used to give an idea for 1 nC and 250 pC in Table \ref{tab:emit_summary}. Their action vs time plots are shown in Fig.~\ref{fig:tail_1nC} and Fig.~\ref{fig:tail_250pC}. Both cases show the large action particles are concentrated in temporal tails, where space charge force is weakest and has a mismatch with solenoid focusing. Large action particles are also distributed near the peak current location, where space charge force is the strongest, and residual nonlinear space charge effect degrades some large radius particles. After flattop laser shaping for a more uniform temporal charge distribution, the space charge mismatch from solenoid focusing near temporal tails is reduced, where the large action particles are also reduced. This is also shown by comparing Fig.~\ref{fig:tail_1nC} and Fig.~\ref{fig:tail_250pC}.

\section{Summary}
\label{summary}
Transverse emittance is an important parameter for optimizing the injector beam brightness, but a large spread of emittance results exists due to its sensitivity to tail particles in the phase space. This makes it difficult to translate emittance to beam brightness and to do absolute comparisons of different injector brightness. An analytical model, consisting of a constant brightness for 95\% charge and a continuous brightness degradation for 5\% charge, demonstrates the 100\% emittance diverges while the beam brightness degradation converges. This explains the 100\% emittance is not a good parameter for characterizing the beam brightness, and an emittance with proper reduced charge is needed. Different ways of cutting non-Gaussian tail particles in phase space or beam transverse profile were used in the past, but several problems exists. For example, the charge cut in transverse beam profile is not self consistent, only an approximation, and the emittance vs charge cut in phase space does not give a final conclusion on beam brightness and percentage of non-Gaussian tails.

The action and phase coordinates are introduced for beam phase space description, which reduces one degree of freedom and makes it more intuitive for the brightness analysis of a Gaussian phase space. In a well optimized photoinjector, the phase space consists of a dominating Gaussian core distribution and a minor non-Gaussian tail distribution, which is clearly observed when projecting the beam intensities to the action axis. The charge and emittance of such a Gaussian mode is defined to be the core charge and core emittance, which is also much less sensitive to the SNR of the measurements. Besides the core emittance, an emittance scaling technique is also introduced to recover the 100\% emittance, the so called scaled emittance can reduce the sensitivity of measured emittance to SNR.

The transverse phase space analysis in the action and phase coordinates is applied to the four bunch charges of the European XFEL injector, but the beam was produced at PITZ by mimicing the European XFEL gun and laser configurations. Measurements in Table \ref{tab:emit_summary} show the charge fraction in the Gaussian core is between 77\% to 91\%, and increases with lower bunch charge. The actual 4D beam brightness in the Gaussian core is 15\% to 57\% higher than that based on the nominal charge and emittance. The 4D beam brightness of the 250 pC beam is about 50\% higher than the original design goal of 1 nC/\textmu m$^2$, and can be further improved to 2 nC/\textmu m$^2$ when charge is reduced to 100 pC. The injector core emittance is dominated by thermal emittance, and the ratio of $\varepsilon_{\rm cath}/\varepsilon_{\rm core}$ is between 63\% to 75\%. Cathode thermal emittance has become the bottleneck of further beam brightness improvement, and R\&D on both Alkali-antimonide cathodes and Cs$_2$Te cathode are ongoing in a collaboration between DESY and LASA at INFN Milano.

\section{Acknowledgements}
The authors appreciate inspiring discussions with C. J. Richard on transverse phase space analysis in the action and phase coordinates. This work was supported by the European XFEL research and development program.
\bibliography{apssamp}% Produces the bibliography via BibTeX.

\providecommand{\noopsort}[1]{}\providecommand{\singleletter}[1]{#1}%
\begin{thebibliography}{40}%
\makeatletter
\providecommand \@ifxundefined [1]{%
 \@ifx{#1\undefined}
}%
\providecommand \@ifnum [1]{%
 \ifnum #1\expandafter \@firstoftwo
 \else \expandafter \@secondoftwo
 \fi
}%
\providecommand \@ifx [1]{%
 \ifx #1\expandafter \@firstoftwo
 \else \expandafter \@secondoftwo
 \fi
}%
\providecommand \natexlab [1]{#1}%
\providecommand \enquote  [1]{``#1''}%
\providecommand \bibnamefont  [1]{#1}%
\providecommand \bibfnamefont [1]{#1}%
\providecommand \citenamefont [1]{#1}%
\providecommand \href@noop [0]{\@secondoftwo}%
\providecommand \href [0]{\begingroup \@sanitize@url \@href}%
\providecommand \@href[1]{\@@startlink{#1}\@@href}%
\providecommand \@@href[1]{\endgroup#1\@@endlink}%
\providecommand \@sanitize@url [0]{\catcode `\\12\catcode `\$12\catcode
  `\&12\catcode `\#12\catcode `\^12\catcode `\_12\catcode `\%12\relax}%
\providecommand \@@startlink[1]{}%
\providecommand \@@endlink[0]{}%
\providecommand \url  [0]{\begingroup\@sanitize@url \@url }%
\providecommand \@url [1]{\endgroup\@href {#1}{\urlprefix }}%
\providecommand \urlprefix  [0]{URL }%
\providecommand \Eprint [0]{\href }%
\providecommand \doibase [0]{https://doi.org/}%
\providecommand \selectlanguage [0]{\@gobble}%
\providecommand \bibinfo  [0]{\@secondoftwo}%
\providecommand \bibfield  [0]{\@secondoftwo}%
\providecommand \translation [1]{[#1]}%
\providecommand \BibitemOpen [0]{}%
\providecommand \bibitemStop [0]{}%
\providecommand \bibitemNoStop [0]{.\EOS\space}%
\providecommand \EOS [0]{\spacefactor3000\relax}%
\providecommand \BibitemShut  [1]{\csname bibitem#1\endcsname}%
\let\auto@bib@innerbib\@empty
%</preamble>
\bibitem [{\citenamefont {Rao}\ and\ \citenamefont
  {Dowell}(2014)}]{rao2014engineering}%
  \BibitemOpen
  \bibfield  {author} {\bibinfo {author} {\bibfnamefont {T.}~\bibnamefont
  {Rao}}\ and\ \bibinfo {author} {\bibfnamefont {D.~H.}\ \bibnamefont
  {Dowell}},\ }\bibfield  {title} {\bibinfo {title} {An engineering guide to
  photoinjectors},\ }\href@noop {} {\bibfield  {journal} {\bibinfo  {journal}
  {arXiv preprint arXiv:1403.7539}\ } (\bibinfo {year} {2014})}\BibitemShut
  {NoStop}%
\bibitem [{\citenamefont {Pellegrini}(2017)}]{pellegrini2017x}%
  \BibitemOpen
  \bibfield  {author} {\bibinfo {author} {\bibfnamefont {C.}~\bibnamefont
  {Pellegrini}},\ }\bibfield  {title} {\bibinfo {title} {X-ray free-electron
  lasers: from dreams to reality},\ }\href@noop {} {\bibfield  {journal}
  {\bibinfo  {journal} {Physica Scripta}\ }\textbf {\bibinfo {volume} {2016}},\
  \bibinfo {pages} {014004} (\bibinfo {year} {2017})}\BibitemShut {NoStop}%
\bibitem [{\citenamefont {Weathersby}\ \emph {et~al.}(2015)\citenamefont
  {Weathersby}, \citenamefont {Brown}, \citenamefont {Centurion}, \citenamefont
  {Chase}, \citenamefont {Coffee}, \citenamefont {Corbett}, \citenamefont
  {Eichner}, \citenamefont {Frisch}, \citenamefont {Fry}, \citenamefont
  {G{\"u}hr} \emph {et~al.}}]{weathersby2015mega}%
  \BibitemOpen
  \bibfield  {author} {\bibinfo {author} {\bibfnamefont {S.}~\bibnamefont
  {Weathersby}}, \bibinfo {author} {\bibfnamefont {G.}~\bibnamefont {Brown}},
  \bibinfo {author} {\bibfnamefont {M.}~\bibnamefont {Centurion}}, \bibinfo
  {author} {\bibfnamefont {T.}~\bibnamefont {Chase}}, \bibinfo {author}
  {\bibfnamefont {R.}~\bibnamefont {Coffee}}, \bibinfo {author} {\bibfnamefont
  {J.}~\bibnamefont {Corbett}}, \bibinfo {author} {\bibfnamefont
  {J.}~\bibnamefont {Eichner}}, \bibinfo {author} {\bibfnamefont
  {J.}~\bibnamefont {Frisch}}, \bibinfo {author} {\bibfnamefont
  {A.}~\bibnamefont {Fry}}, \bibinfo {author} {\bibfnamefont {M.}~\bibnamefont
  {G{\"u}hr}}, \emph {et~al.},\ }\bibfield  {title} {\bibinfo {title}
  {Mega-electron-volt ultrafast electron diffraction at slac national
  accelerator laboratory},\ }\href@noop {} {\bibfield  {journal} {\bibinfo
  {journal} {Review of Scientific Instruments}\ }\textbf {\bibinfo {volume}
  {86}},\ \bibinfo {pages} {073702} (\bibinfo {year} {2015})}\BibitemShut
  {NoStop}%
\bibitem [{\citenamefont {Merminga}(2020)}]{merminga2020energy}%
  \BibitemOpen
  \bibfield  {author} {\bibinfo {author} {\bibfnamefont {L.}~\bibnamefont
  {Merminga}},\ }\bibfield  {title} {\bibinfo {title} {Energy recovery
  linacs},\ }\href@noop {} {\bibfield  {journal} {\bibinfo  {journal}
  {Synchrotron Light Sources and Free-Electron Lasers: Accelerator Physics,
  Instrumentation and Science Applications}\ ,\ \bibinfo {pages} {439}}
  (\bibinfo {year} {2020})}\BibitemShut {NoStop}%
\bibitem [{\citenamefont {Huang}\ and\ \citenamefont
  {Kim}(2007)}]{huang2007review}%
  \BibitemOpen
  \bibfield  {author} {\bibinfo {author} {\bibfnamefont {Z.}~\bibnamefont
  {Huang}}\ and\ \bibinfo {author} {\bibfnamefont {K.-J.}\ \bibnamefont
  {Kim}},\ }\bibfield  {title} {\bibinfo {title} {Review of x-ray free-electron
  laser theory},\ }\href@noop {} {\bibfield  {journal} {\bibinfo  {journal}
  {Physical Review Special Topics-Accelerators and Beams}\ }\textbf {\bibinfo
  {volume} {10}},\ \bibinfo {pages} {034801} (\bibinfo {year}
  {2007})}\BibitemShut {NoStop}%
\bibitem [{\citenamefont {Nuhn}(2002)}]{nuhn2002linac}%
  \BibitemOpen
  \bibfield  {author} {\bibinfo {author} {\bibfnamefont {H.-D.}\ \bibnamefont
  {Nuhn}},\ }\href@noop {} {\emph {\bibinfo {title} {Linac Coherent Light
  Source (LCLS) Conceptual Design Report}}},\ \bibinfo {type} {Tech. Rep.}\
  (\bibinfo  {institution} {Stanford Linear Accelerator Center},\ \bibinfo
  {year} {2002})\BibitemShut {NoStop}%
\bibitem [{\citenamefont {Decking}\ and\ \citenamefont
  {Limberg}(2013)}]{decking2013european}%
  \BibitemOpen
  \bibfield  {author} {\bibinfo {author} {\bibfnamefont {W.}~\bibnamefont
  {Decking}}\ and\ \bibinfo {author} {\bibfnamefont {T.}~\bibnamefont
  {Limberg}},\ }\bibfield  {title} {\bibinfo {title} {European xfel post-tdr
  description},\ }\href@noop {} {\bibfield  {journal} {\bibinfo  {journal}
  {XFEL. EU TN-2013-004-01, European XFEL GmbH, Hamburg, Germany}\ } (\bibinfo
  {year} {2013})}\BibitemShut {NoStop}%
\bibitem [{\citenamefont {Emma}\ \emph {et~al.}(2010)\citenamefont {Emma},
  \citenamefont {Akre}, \citenamefont {Arthur}, \citenamefont {Bionta},
  \citenamefont {Bostedt}, \citenamefont {Bozek}, \citenamefont {Brachmann},
  \citenamefont {Bucksbaum}, \citenamefont {Coffee}, \citenamefont {Decker}
  \emph {et~al.}}]{emma2010first}%
  \BibitemOpen
  \bibfield  {author} {\bibinfo {author} {\bibfnamefont {P.}~\bibnamefont
  {Emma}}, \bibinfo {author} {\bibfnamefont {R.}~\bibnamefont {Akre}}, \bibinfo
  {author} {\bibfnamefont {J.}~\bibnamefont {Arthur}}, \bibinfo {author}
  {\bibfnamefont {R.}~\bibnamefont {Bionta}}, \bibinfo {author} {\bibfnamefont
  {C.}~\bibnamefont {Bostedt}}, \bibinfo {author} {\bibfnamefont
  {J.}~\bibnamefont {Bozek}}, \bibinfo {author} {\bibfnamefont
  {A.}~\bibnamefont {Brachmann}}, \bibinfo {author} {\bibfnamefont
  {P.}~\bibnamefont {Bucksbaum}}, \bibinfo {author} {\bibfnamefont
  {R.}~\bibnamefont {Coffee}}, \bibinfo {author} {\bibfnamefont {F.-J.}\
  \bibnamefont {Decker}}, \emph {et~al.},\ }\bibfield  {title} {\bibinfo
  {title} {First lasing and operation of an {\aa}ngstrom-wavelength
  free-electron laser},\ }\href@noop {} {\bibfield  {journal} {\bibinfo
  {journal} {nature photonics}\ }\textbf {\bibinfo {volume} {4}},\ \bibinfo
  {pages} {641} (\bibinfo {year} {2010})}\BibitemShut {NoStop}%
\bibitem [{\citenamefont {Prat}\ \emph {et~al.}(2020)\citenamefont {Prat},
  \citenamefont {Abela}, \citenamefont {Aiba}, \citenamefont {Alarcon},
  \citenamefont {Alex}, \citenamefont {Arbelo}, \citenamefont {Arrell},
  \citenamefont {Arsov}, \citenamefont {Bacellar}, \citenamefont {Beard} \emph
  {et~al.}}]{prat2020compact}%
  \BibitemOpen
  \bibfield  {author} {\bibinfo {author} {\bibfnamefont {E.}~\bibnamefont
  {Prat}}, \bibinfo {author} {\bibfnamefont {R.}~\bibnamefont {Abela}},
  \bibinfo {author} {\bibfnamefont {M.}~\bibnamefont {Aiba}}, \bibinfo {author}
  {\bibfnamefont {A.}~\bibnamefont {Alarcon}}, \bibinfo {author} {\bibfnamefont
  {J.}~\bibnamefont {Alex}}, \bibinfo {author} {\bibfnamefont {Y.}~\bibnamefont
  {Arbelo}}, \bibinfo {author} {\bibfnamefont {C.}~\bibnamefont {Arrell}},
  \bibinfo {author} {\bibfnamefont {V.}~\bibnamefont {Arsov}}, \bibinfo
  {author} {\bibfnamefont {C.}~\bibnamefont {Bacellar}}, \bibinfo {author}
  {\bibfnamefont {C.}~\bibnamefont {Beard}}, \emph {et~al.},\ }\bibfield
  {title} {\bibinfo {title} {A compact and cost-effective hard x-ray
  free-electron laser driven by a high-brightness and low-energy electron
  beam},\ }\href@noop {} {\bibfield  {journal} {\bibinfo  {journal} {Nature
  Photonics}\ }\textbf {\bibinfo {volume} {14}},\ \bibinfo {pages} {748}
  (\bibinfo {year} {2020})}\BibitemShut {NoStop}%
\bibitem [{\citenamefont {Decking}\ \emph {et~al.}(2020)\citenamefont
  {Decking}, \citenamefont {Abeghyan}, \citenamefont {Abramian}, \citenamefont
  {Abramsky}, \citenamefont {Aguirre}, \citenamefont {Albrecht}, \citenamefont
  {Alou}, \citenamefont {Altarelli}, \citenamefont {Altmann}, \citenamefont
  {Amyan} \emph {et~al.}}]{decking2020mhz}%
  \BibitemOpen
  \bibfield  {author} {\bibinfo {author} {\bibfnamefont {W.}~\bibnamefont
  {Decking}}, \bibinfo {author} {\bibfnamefont {S.}~\bibnamefont {Abeghyan}},
  \bibinfo {author} {\bibfnamefont {P.}~\bibnamefont {Abramian}}, \bibinfo
  {author} {\bibfnamefont {A.}~\bibnamefont {Abramsky}}, \bibinfo {author}
  {\bibfnamefont {A.}~\bibnamefont {Aguirre}}, \bibinfo {author} {\bibfnamefont
  {C.}~\bibnamefont {Albrecht}}, \bibinfo {author} {\bibfnamefont
  {P.}~\bibnamefont {Alou}}, \bibinfo {author} {\bibfnamefont {M.}~\bibnamefont
  {Altarelli}}, \bibinfo {author} {\bibfnamefont {P.}~\bibnamefont {Altmann}},
  \bibinfo {author} {\bibfnamefont {K.}~\bibnamefont {Amyan}}, \emph {et~al.},\
  }\bibfield  {title} {\bibinfo {title} {A mhz-repetition-rate hard x-ray
  free-electron laser driven by a superconducting linear accelerator},\
  }\href@noop {} {\bibfield  {journal} {\bibinfo  {journal} {Nature Photonics}\
  }\textbf {\bibinfo {volume} {14}},\ \bibinfo {pages} {391} (\bibinfo {year}
  {2020})}\BibitemShut {NoStop}%
\bibitem [{\citenamefont {Rosenzweig}\ \emph {et~al.}(2020)\citenamefont
  {Rosenzweig}, \citenamefont {Majernik}, \citenamefont {Robles}, \citenamefont
  {Andonian}, \citenamefont {Camacho}, \citenamefont {Fukasawa}, \citenamefont
  {Kogar}, \citenamefont {Lawler}, \citenamefont {Miao}, \citenamefont
  {Musumeci} \emph {et~al.}}]{rosenzweig2020ultra}%
  \BibitemOpen
  \bibfield  {author} {\bibinfo {author} {\bibfnamefont {J.}~\bibnamefont
  {Rosenzweig}}, \bibinfo {author} {\bibfnamefont {N.}~\bibnamefont
  {Majernik}}, \bibinfo {author} {\bibfnamefont {R.}~\bibnamefont {Robles}},
  \bibinfo {author} {\bibfnamefont {G.}~\bibnamefont {Andonian}}, \bibinfo
  {author} {\bibfnamefont {O.}~\bibnamefont {Camacho}}, \bibinfo {author}
  {\bibfnamefont {A.}~\bibnamefont {Fukasawa}}, \bibinfo {author}
  {\bibfnamefont {A.}~\bibnamefont {Kogar}}, \bibinfo {author} {\bibfnamefont
  {G.}~\bibnamefont {Lawler}}, \bibinfo {author} {\bibfnamefont
  {J.}~\bibnamefont {Miao}}, \bibinfo {author} {\bibfnamefont {P.}~\bibnamefont
  {Musumeci}}, \emph {et~al.},\ }\bibfield  {title} {\bibinfo {title} {An
  ultra-compact x-ray free-electron laser},\ }\href@noop {} {\bibfield
  {journal} {\bibinfo  {journal} {New Journal of Physics}\ }\textbf {\bibinfo
  {volume} {22}},\ \bibinfo {pages} {093067} (\bibinfo {year}
  {2020})}\BibitemShut {NoStop}%
\bibitem [{\citenamefont {D'Auria}\ \emph {et~al.}(2019)\citenamefont {D'Auria}
  \emph {et~al.}}]{d2019compactlight}%
  \BibitemOpen
  \bibfield  {author} {\bibinfo {author} {\bibfnamefont {G.}~\bibnamefont
  {D'Auria}} \emph {et~al.},\ }\bibfield  {title} {\bibinfo {title} {{T}he
  {C}ompact{L}ight {D}esign {S}tudy {P}roject},\ }in\ \href
  {https://doi.org/doi:10.18429/JACoW-IPAC2019-TUPRB032} {\emph {\bibinfo
  {booktitle} {Proc. 10th International Particle Accelerator Conference
  (IPAC'19), Melbourne, Australia, 19-24 May 2019}}},\ \bibinfo {series and
  number} {\bibinfo {series} {International Particle Accelerator Conference}\
  No.~\bibinfo {number} {10}}\ (\bibinfo  {publisher} {JACoW Publishing},\
  \bibinfo {address} {Geneva, Switzerland},\ \bibinfo {year} {2019})\ pp.\
  \bibinfo {pages} {1756--1759},\ \bibinfo {note}
  {https://doi.org/10.18429/JACoW-IPAC2019-TUPRB032}\BibitemShut {NoStop}%
\bibitem [{\citenamefont {Raubenheimer}\ \emph {et~al.}(2018)\citenamefont
  {Raubenheimer} \emph {et~al.}}]{raubenheimer2018lcls}%
  \BibitemOpen
  \bibfield  {author} {\bibinfo {author} {\bibfnamefont {T.}~\bibnamefont
  {Raubenheimer}} \emph {et~al.},\ }\bibfield  {title} {\bibinfo {title} {The
  lcls-ii-he, a high energy upgrade of the lcls-ii},\ }in\ \href@noop {} {\emph
  {\bibinfo {booktitle} {60th ICFA Advanced Beam Dynamics Workshop on Future
  Light Sources}}}\ (\bibinfo {year} {2018})\ pp.\ \bibinfo {pages}
  {6--11}\BibitemShut {NoStop}%
\bibitem [{\citenamefont {Carlsten}(1995)}]{carlsten1995space}%
  \BibitemOpen
  \bibfield  {author} {\bibinfo {author} {\bibfnamefont {B.~E.}\ \bibnamefont
  {Carlsten}},\ }\bibfield  {title} {\bibinfo {title} {Space charge induced
  emittance compensation in high brightness photoinjectors},\ }\href@noop {}
  {\bibfield  {journal} {\bibinfo  {journal} {Part. Accel.}\ }\textbf {\bibinfo
  {volume} {49}},\ \bibinfo {pages} {27} (\bibinfo {year} {1995})}\BibitemShut
  {NoStop}%
\bibitem [{200(2008)}]{2008emit_workshop}%
  \BibitemOpen
  \href@noop {} {\bibinfo {title} {Mini workshop on `characterization of high
  brightness beams'}} (\bibinfo {year} {2008}),\ \bibinfo {note}
  {\url{https://indico.desy.de/event/806/}}\BibitemShut {NoStop}%
\bibitem [{\citenamefont {Richard}\ \emph {et~al.}(2020)\citenamefont
  {Richard}, \citenamefont {Alvarez}, \citenamefont {Carneiro}, \citenamefont
  {Hanna}, \citenamefont {Prost}, \citenamefont {Saini}, \citenamefont
  {Scarpine},\ and\ \citenamefont {Shemyakin}}]{richard2020measurements}%
  \BibitemOpen
  \bibfield  {author} {\bibinfo {author} {\bibfnamefont {C.}~\bibnamefont
  {Richard}}, \bibinfo {author} {\bibfnamefont {M.}~\bibnamefont {Alvarez}},
  \bibinfo {author} {\bibfnamefont {J.}~\bibnamefont {Carneiro}}, \bibinfo
  {author} {\bibfnamefont {B.}~\bibnamefont {Hanna}}, \bibinfo {author}
  {\bibfnamefont {L.}~\bibnamefont {Prost}}, \bibinfo {author} {\bibfnamefont
  {A.}~\bibnamefont {Saini}}, \bibinfo {author} {\bibfnamefont
  {V.}~\bibnamefont {Scarpine}},\ and\ \bibinfo {author} {\bibfnamefont
  {A.}~\bibnamefont {Shemyakin}},\ }\bibfield  {title} {\bibinfo {title}
  {Measurements of a 2.1 mev h- beam with an allison scanner},\ }\href@noop {}
  {\bibfield  {journal} {\bibinfo  {journal} {Review of Scientific
  Instruments}\ }\textbf {\bibinfo {volume} {91}},\ \bibinfo {pages} {073301}
  (\bibinfo {year} {2020})}\BibitemShut {NoStop}%
\bibitem [{\citenamefont {Krasilnikov}\ \emph {et~al.}(2012)\citenamefont
  {Krasilnikov}, \citenamefont {Stephan}, \citenamefont {Asova}, \citenamefont
  {Grabosch}, \citenamefont {Gro{\ss}}, \citenamefont {Hakobyan}, \citenamefont
  {Isaev}, \citenamefont {Ivanisenko}, \citenamefont {Jachmann}, \citenamefont
  {Khojoyan} \emph {et~al.}}]{MK2012}%
  \BibitemOpen
  \bibfield  {author} {\bibinfo {author} {\bibfnamefont {M.}~\bibnamefont
  {Krasilnikov}}, \bibinfo {author} {\bibfnamefont {F.}~\bibnamefont
  {Stephan}}, \bibinfo {author} {\bibfnamefont {G.}~\bibnamefont {Asova}},
  \bibinfo {author} {\bibfnamefont {H.-J.}\ \bibnamefont {Grabosch}}, \bibinfo
  {author} {\bibfnamefont {M.}~\bibnamefont {Gro{\ss}}}, \bibinfo {author}
  {\bibfnamefont {L.}~\bibnamefont {Hakobyan}}, \bibinfo {author}
  {\bibfnamefont {I.}~\bibnamefont {Isaev}}, \bibinfo {author} {\bibfnamefont
  {Y.}~\bibnamefont {Ivanisenko}}, \bibinfo {author} {\bibfnamefont
  {L.}~\bibnamefont {Jachmann}}, \bibinfo {author} {\bibfnamefont
  {M.}~\bibnamefont {Khojoyan}}, \emph {et~al.},\ }\bibfield  {title} {\bibinfo
  {title} {Experimentally minimized beam emittance from an l-band
  photoinjector},\ }\href@noop {} {\bibfield  {journal} {\bibinfo  {journal}
  {Physical Review Special Topics-Accelerators and Beams}\ }\textbf {\bibinfo
  {volume} {15}},\ \bibinfo {pages} {100701} (\bibinfo {year}
  {2012})}\BibitemShut {NoStop}%
\bibitem [{\citenamefont {Gulliford}\ \emph {et~al.}(2013)\citenamefont
  {Gulliford}, \citenamefont {Bartnik}, \citenamefont {Bazarov}, \citenamefont
  {Cultrera}, \citenamefont {Dobbins}, \citenamefont {Dunham}, \citenamefont
  {Gonzalez}, \citenamefont {Karkare}, \citenamefont {Lee}, \citenamefont {Li}
  \emph {et~al.}}]{gulliford2013demonstration}%
  \BibitemOpen
  \bibfield  {author} {\bibinfo {author} {\bibfnamefont {C.}~\bibnamefont
  {Gulliford}}, \bibinfo {author} {\bibfnamefont {A.}~\bibnamefont {Bartnik}},
  \bibinfo {author} {\bibfnamefont {I.}~\bibnamefont {Bazarov}}, \bibinfo
  {author} {\bibfnamefont {L.}~\bibnamefont {Cultrera}}, \bibinfo {author}
  {\bibfnamefont {J.}~\bibnamefont {Dobbins}}, \bibinfo {author} {\bibfnamefont
  {B.}~\bibnamefont {Dunham}}, \bibinfo {author} {\bibfnamefont
  {F.}~\bibnamefont {Gonzalez}}, \bibinfo {author} {\bibfnamefont
  {S.}~\bibnamefont {Karkare}}, \bibinfo {author} {\bibfnamefont
  {H.}~\bibnamefont {Lee}}, \bibinfo {author} {\bibfnamefont {H.}~\bibnamefont
  {Li}}, \emph {et~al.},\ }\bibfield  {title} {\bibinfo {title} {Demonstration
  of low emittance in the cornell energy recovery linac injector prototype},\
  }\href@noop {} {\bibfield  {journal} {\bibinfo  {journal} {Physical Review
  Special Topics-Accelerators and Beams}\ }\textbf {\bibinfo {volume} {16}},\
  \bibinfo {pages} {073401} (\bibinfo {year} {2013})}\BibitemShut {NoStop}%
\bibitem [{\citenamefont {L{\"o}hl}(2005)}]{lohl2005measurements}%
  \BibitemOpen
  \bibfield  {author} {\bibinfo {author} {\bibfnamefont {F.}~\bibnamefont
  {L{\"o}hl}},\ }\href@noop {} {\emph {\bibinfo {title} {Measurements of the
  Transverse Emittance at the VUV-FEL}}},\ \bibinfo {type} {Tech. Rep.}\
  (\bibinfo  {institution} {DESY},\ \bibinfo {year} {2005})\BibitemShut
  {NoStop}%
\bibitem [{\citenamefont {Prat}\ \emph
  {et~al.}(2014{\natexlab{a}})\citenamefont {Prat}, \citenamefont {Aiba},
  \citenamefont {Bettoni}, \citenamefont {Beutner}, \citenamefont {Reiche},\
  and\ \citenamefont {Schietinger}}]{prat2014emittance}%
  \BibitemOpen
  \bibfield  {author} {\bibinfo {author} {\bibfnamefont {E.}~\bibnamefont
  {Prat}}, \bibinfo {author} {\bibfnamefont {M.}~\bibnamefont {Aiba}}, \bibinfo
  {author} {\bibfnamefont {S.}~\bibnamefont {Bettoni}}, \bibinfo {author}
  {\bibfnamefont {B.}~\bibnamefont {Beutner}}, \bibinfo {author} {\bibfnamefont
  {S.}~\bibnamefont {Reiche}},\ and\ \bibinfo {author} {\bibfnamefont
  {T.}~\bibnamefont {Schietinger}},\ }\bibfield  {title} {\bibinfo {title}
  {Emittance measurements and minimization at the swissfel injector test
  facility},\ }\href@noop {} {\bibfield  {journal} {\bibinfo  {journal}
  {Physical Review Special Topics-Accelerators and Beams}\ }\textbf {\bibinfo
  {volume} {17}},\ \bibinfo {pages} {104401} (\bibinfo {year}
  {2014}{\natexlab{a}})}\BibitemShut {NoStop}%
\bibitem [{\citenamefont {Stupakov}\ and\ \citenamefont
  {Penn}(2018)}]{stupakov2018classical}%
  \BibitemOpen
  \bibfield  {author} {\bibinfo {author} {\bibfnamefont {G.}~\bibnamefont
  {Stupakov}}\ and\ \bibinfo {author} {\bibfnamefont {G.}~\bibnamefont
  {Penn}},\ }\href@noop {} {\emph {\bibinfo {title} {Classical mechanics and
  electromagnetism in accelerator physics}}},\ Vol.~\bibinfo {volume} {61}\
  (\bibinfo  {publisher} {Springer},\ \bibinfo {year} {2018})\BibitemShut
  {NoStop}%
\bibitem [{\citenamefont {Bazarov}\ \emph {et~al.}(2009)\citenamefont
  {Bazarov}, \citenamefont {Dunham},\ and\ \citenamefont
  {Sinclair}}]{bazarov2009maximum}%
  \BibitemOpen
  \bibfield  {author} {\bibinfo {author} {\bibfnamefont {I.~V.}\ \bibnamefont
  {Bazarov}}, \bibinfo {author} {\bibfnamefont {B.~M.}\ \bibnamefont
  {Dunham}},\ and\ \bibinfo {author} {\bibfnamefont {C.~K.}\ \bibnamefont
  {Sinclair}},\ }\bibfield  {title} {\bibinfo {title} {Maximum achievable beam
  brightness from photoinjectors},\ }\href@noop {} {\bibfield  {journal}
  {\bibinfo  {journal} {Physical review letters}\ }\textbf {\bibinfo {volume}
  {102}},\ \bibinfo {pages} {104801} (\bibinfo {year} {2009})}\BibitemShut
  {NoStop}%
\bibitem [{\citenamefont {Stephan}\ \emph {et~al.}(2010)\citenamefont
  {Stephan}, \citenamefont {Boulware}, \citenamefont {Krasilnikov},
  \citenamefont {B{\"a}hr}, \citenamefont {Asova}, \citenamefont {Donat},
  \citenamefont {Gensch}, \citenamefont {Grabosch}, \citenamefont {H{\"a}nel},
  \citenamefont {Hakobyan} \emph {et~al.}}]{FS2010}%
  \BibitemOpen
  \bibfield  {author} {\bibinfo {author} {\bibfnamefont {F.}~\bibnamefont
  {Stephan}}, \bibinfo {author} {\bibfnamefont {C.}~\bibnamefont {Boulware}},
  \bibinfo {author} {\bibfnamefont {M.}~\bibnamefont {Krasilnikov}}, \bibinfo
  {author} {\bibfnamefont {J.}~\bibnamefont {B{\"a}hr}}, \bibinfo {author}
  {\bibfnamefont {G.}~\bibnamefont {Asova}}, \bibinfo {author} {\bibfnamefont
  {A.}~\bibnamefont {Donat}}, \bibinfo {author} {\bibfnamefont
  {U.}~\bibnamefont {Gensch}}, \bibinfo {author} {\bibfnamefont
  {H.}~\bibnamefont {Grabosch}}, \bibinfo {author} {\bibfnamefont
  {M.}~\bibnamefont {H{\"a}nel}}, \bibinfo {author} {\bibfnamefont
  {L.}~\bibnamefont {Hakobyan}}, \emph {et~al.},\ }\bibfield  {title} {\bibinfo
  {title} {Detailed characterization of electron sources yielding first
  demonstration of european x-ray free-electron laser beam quality},\
  }\href@noop {} {\bibfield  {journal} {\bibinfo  {journal} {Physical Review
  Special Topics-Accelerators and Beams}\ }\textbf {\bibinfo {volume} {13}},\
  \bibinfo {pages} {020704} (\bibinfo {year} {2010})}\BibitemShut {NoStop}%
\bibitem [{\citenamefont {Dwersteg}\ \emph {et~al.}(1997)\citenamefont
  {Dwersteg}, \citenamefont {Fl{\"o}ttmann}, \citenamefont {Sekutowicz},\ and\
  \citenamefont {Stolzenburg}}]{BD1997}%
  \BibitemOpen
  \bibfield  {author} {\bibinfo {author} {\bibfnamefont {B.}~\bibnamefont
  {Dwersteg}}, \bibinfo {author} {\bibfnamefont {K.}~\bibnamefont
  {Fl{\"o}ttmann}}, \bibinfo {author} {\bibfnamefont {J.}~\bibnamefont
  {Sekutowicz}},\ and\ \bibinfo {author} {\bibfnamefont {C.}~\bibnamefont
  {Stolzenburg}},\ }\bibfield  {title} {\bibinfo {title} {Rf gun design for the
  tesla vuv free electron laser},\ }\href@noop {} {\bibfield  {journal}
  {\bibinfo  {journal} {Nuclear Instruments and Methods in Physics Research
  Section A: Accelerators, Spectrometers, Detectors and Associated Equipment}\
  }\textbf {\bibinfo {volume} {393}},\ \bibinfo {pages} {93} (\bibinfo {year}
  {1997})}\BibitemShut {NoStop}%
\bibitem [{\citenamefont {Paramonov}\ \emph {et~al.}(2017)\citenamefont
  {Paramonov}, \citenamefont {Philipp}, \citenamefont {Rybakov}, \citenamefont
  {Skassyrskaya},\ and\ \citenamefont {Stephan}}]{paramonov2017design}%
  \BibitemOpen
  \bibfield  {author} {\bibinfo {author} {\bibfnamefont {V.}~\bibnamefont
  {Paramonov}}, \bibinfo {author} {\bibfnamefont {S.}~\bibnamefont {Philipp}},
  \bibinfo {author} {\bibfnamefont {I.}~\bibnamefont {Rybakov}}, \bibinfo
  {author} {\bibfnamefont {A.}~\bibnamefont {Skassyrskaya}},\ and\ \bibinfo
  {author} {\bibfnamefont {F.}~\bibnamefont {Stephan}},\ }\bibfield  {title}
  {\bibinfo {title} {Design of an l-band normally conducting rf gun cavity for
  high peak and average rf power},\ }\href@noop {} {\bibfield  {journal}
  {\bibinfo  {journal} {Nuclear Instruments and Methods in Physics Research
  Section A: Accelerators, Spectrometers, Detectors and Associated Equipment}\
  }\textbf {\bibinfo {volume} {854}},\ \bibinfo {pages} {113} (\bibinfo {year}
  {2017})}\BibitemShut {NoStop}%
\bibitem [{\citenamefont {Lederer}\ \emph {et~al.}(2018)\citenamefont
  {Lederer}, \citenamefont {Schreiber} \emph {et~al.}}]{lederer2018cs2te}%
  \BibitemOpen
  \bibfield  {author} {\bibinfo {author} {\bibfnamefont {S.}~\bibnamefont
  {Lederer}}, \bibinfo {author} {\bibfnamefont {S.}~\bibnamefont {Schreiber}},
  \emph {et~al.},\ }\bibfield  {title} {\bibinfo {title} {Cs2te photocathode
  lifetime at {FLASH} and {E}uropean {XFEL}},\ }\href@noop {} {\bibfield
  {journal} {\bibinfo  {journal} {Proc. IPAC’18}\ ,\ \bibinfo {pages} {2496}}
  (\bibinfo {year} {2018})}\BibitemShut {NoStop}%
\bibitem [{\citenamefont {Huang}\ \emph {et~al.}(2020)\citenamefont {Huang},
  \citenamefont {Qian}, \citenamefont {Chen}, \citenamefont {Filippetto},
  \citenamefont {Gross}, \citenamefont {Isaev}, \citenamefont {Koschitzki},
  \citenamefont {Krasilnikov}, \citenamefont {Lal}, \citenamefont {Li} \emph
  {et~al.}}]{huang2020single}%
  \BibitemOpen
  \bibfield  {author} {\bibinfo {author} {\bibfnamefont {P.-W.}\ \bibnamefont
  {Huang}}, \bibinfo {author} {\bibfnamefont {H.}~\bibnamefont {Qian}},
  \bibinfo {author} {\bibfnamefont {Y.}~\bibnamefont {Chen}}, \bibinfo {author}
  {\bibfnamefont {D.}~\bibnamefont {Filippetto}}, \bibinfo {author}
  {\bibfnamefont {M.}~\bibnamefont {Gross}}, \bibinfo {author} {\bibfnamefont
  {I.}~\bibnamefont {Isaev}}, \bibinfo {author} {\bibfnamefont
  {C.}~\bibnamefont {Koschitzki}}, \bibinfo {author} {\bibfnamefont
  {M.}~\bibnamefont {Krasilnikov}}, \bibinfo {author} {\bibfnamefont
  {S.}~\bibnamefont {Lal}}, \bibinfo {author} {\bibfnamefont {X.}~\bibnamefont
  {Li}}, \emph {et~al.},\ }\bibfield  {title} {\bibinfo {title} {Single shot
  cathode transverse momentum imaging in high brightness photoinjectors},\
  }\href@noop {} {\bibfield  {journal} {\bibinfo  {journal} {Physical Review
  Accelerators and Beams}\ }\textbf {\bibinfo {volume} {23}},\ \bibinfo {pages}
  {043401} (\bibinfo {year} {2020})}\BibitemShut {NoStop}%
\bibitem [{\citenamefont {Huang}\ \emph {et~al.}(2019)\citenamefont {Huang},
  \citenamefont {Qian}, \citenamefont {Chen}, \citenamefont {Grigoryan},
  \citenamefont {Gross}, \citenamefont {Isaev}, \citenamefont {Kitisri},
  \citenamefont {Koschitzki}, \citenamefont {Krasilnikov}, \citenamefont {Lal}
  \emph {et~al.}}]{huang2019test}%
  \BibitemOpen
  \bibfield  {author} {\bibinfo {author} {\bibfnamefont {P.}~\bibnamefont
  {Huang}}, \bibinfo {author} {\bibfnamefont {H.}~\bibnamefont {Qian}},
  \bibinfo {author} {\bibfnamefont {Y.}~\bibnamefont {Chen}}, \bibinfo {author}
  {\bibfnamefont {A.}~\bibnamefont {Grigoryan}}, \bibinfo {author}
  {\bibfnamefont {M.}~\bibnamefont {Gross}}, \bibinfo {author} {\bibfnamefont
  {I.}~\bibnamefont {Isaev}}, \bibinfo {author} {\bibfnamefont
  {P.}~\bibnamefont {Kitisri}}, \bibinfo {author} {\bibfnamefont
  {C.}~\bibnamefont {Koschitzki}}, \bibinfo {author} {\bibfnamefont
  {M.}~\bibnamefont {Krasilnikov}}, \bibinfo {author} {\bibfnamefont
  {S.}~\bibnamefont {Lal}}, \emph {et~al.},\ }\bibfield  {title} {\bibinfo
  {title} {Test of cs2te thickness on cathode performance at pitz},\ }in\
  \href@noop {} {\emph {\bibinfo {booktitle} {39th Int. Free Electron Laser
  Conf.(FEL'19), Hamburg, Germany}}}\ (\bibinfo {year} {2019})\BibitemShut
  {NoStop}%
\bibitem [{\citenamefont {Will}\ and\ \citenamefont
  {Klemz}(2008)}]{will2008generation}%
  \BibitemOpen
  \bibfield  {author} {\bibinfo {author} {\bibfnamefont {I.}~\bibnamefont
  {Will}}\ and\ \bibinfo {author} {\bibfnamefont {G.}~\bibnamefont {Klemz}},\
  }\bibfield  {title} {\bibinfo {title} {Generation of flat-top picosecond
  pulses by coherent pulse stacking in a multicrystal birefringent filter},\
  }\href@noop {} {\bibfield  {journal} {\bibinfo  {journal} {Optics express}\
  }\textbf {\bibinfo {volume} {16}},\ \bibinfo {pages} {14922} (\bibinfo {year}
  {2008})}\BibitemShut {NoStop}%
\bibitem [{\citenamefont {Gross}\ \emph {et~al.}(2019)\citenamefont {Gross},
  \citenamefont {Qian}, \citenamefont {Boonpornprasert}, \citenamefont {Chen},
  \citenamefont {Good}, \citenamefont {Huck}, \citenamefont {Isaev},
  \citenamefont {Koschitzki}, \citenamefont {Krasilnikov}, \citenamefont {Lal}
  \emph {et~al.}}]{gross2019emittance}%
  \BibitemOpen
  \bibfield  {author} {\bibinfo {author} {\bibfnamefont {M.}~\bibnamefont
  {Gross}}, \bibinfo {author} {\bibfnamefont {H.}~\bibnamefont {Qian}},
  \bibinfo {author} {\bibfnamefont {P.}~\bibnamefont {Boonpornprasert}},
  \bibinfo {author} {\bibfnamefont {Y.}~\bibnamefont {Chen}}, \bibinfo {author}
  {\bibfnamefont {J.}~\bibnamefont {Good}}, \bibinfo {author} {\bibfnamefont
  {H.}~\bibnamefont {Huck}}, \bibinfo {author} {\bibfnamefont {I.}~\bibnamefont
  {Isaev}}, \bibinfo {author} {\bibfnamefont {C.}~\bibnamefont {Koschitzki}},
  \bibinfo {author} {\bibfnamefont {M.}~\bibnamefont {Krasilnikov}}, \bibinfo
  {author} {\bibfnamefont {S.}~\bibnamefont {Lal}}, \emph {et~al.},\ }\bibfield
   {title} {\bibinfo {title} {Emittance reduction of rf photoinjector generated
  electron beams by transverse laser beam shaping},\ }in\ \href@noop {} {\emph
  {\bibinfo {booktitle} {Journal of Physics: Conference Series}}},\ Vol.\
  \bibinfo {volume} {1350}\ (\bibinfo {organization} {IOP Publishing},\
  \bibinfo {year} {2019})\ p.\ \bibinfo {pages} {012046}\BibitemShut {NoStop}%
\bibitem [{\citenamefont {Staykov}\ and\ \citenamefont
  {Tsakov}(2005)}]{staykov2005design}%
  \BibitemOpen
  \bibfield  {author} {\bibinfo {author} {\bibfnamefont {L.}~\bibnamefont
  {Staykov}}\ and\ \bibinfo {author} {\bibfnamefont {I.}~\bibnamefont
  {Tsakov}},\ }\bibfield  {title} {\bibinfo {title} {Design optimization of an
  emittance measurement system at pitz},\ }\href@noop {} {\bibfield  {journal}
  {\bibinfo  {journal} {proceedings DIPAC 2005}\ } (\bibinfo {year}
  {2005})}\BibitemShut {NoStop}%
\bibitem [{\citenamefont {Niemczyk}(2021)}]{niemczyk2021subpicosecond}%
  \BibitemOpen
  \bibfield  {author} {\bibinfo {author} {\bibfnamefont {R.}~\bibnamefont
  {Niemczyk}},\ }\emph {\bibinfo {title} {Subpicosecond-resolved emittance
  measurements of high-brightness electron beams with space charge effects at
  PITZ}},\ \href@noop {} {\bibinfo {type} {{PhD} dissertation}},\ \bibinfo
  {school} {Technische Universität Hamburg} (\bibinfo {year}
  {2021})\BibitemShut {NoStop}%
\bibitem [{\citenamefont {Rimjaem}\ \emph {et~al.}(2010)\citenamefont
  {Rimjaem}, \citenamefont {Asova}, \citenamefont {B{\"a}hr}, \citenamefont
  {Grabosch}, \citenamefont {H{\"a}nel}, \citenamefont {Hakobyan},
  \citenamefont {Ivanisenko}, \citenamefont {Khojoyan}, \citenamefont {Klemz},
  \citenamefont {Krasilnikov} \emph {et~al.}}]{rimjaem2010generating}%
  \BibitemOpen
  \bibfield  {author} {\bibinfo {author} {\bibfnamefont {S.}~\bibnamefont
  {Rimjaem}}, \bibinfo {author} {\bibfnamefont {G.}~\bibnamefont {Asova}},
  \bibinfo {author} {\bibfnamefont {J.}~\bibnamefont {B{\"a}hr}}, \bibinfo
  {author} {\bibfnamefont {H.}~\bibnamefont {Grabosch}}, \bibinfo {author}
  {\bibfnamefont {M.}~\bibnamefont {H{\"a}nel}}, \bibinfo {author}
  {\bibfnamefont {L.}~\bibnamefont {Hakobyan}}, \bibinfo {author}
  {\bibfnamefont {Y.}~\bibnamefont {Ivanisenko}}, \bibinfo {author}
  {\bibfnamefont {M.}~\bibnamefont {Khojoyan}}, \bibinfo {author}
  {\bibfnamefont {G.}~\bibnamefont {Klemz}}, \bibinfo {author} {\bibfnamefont
  {M.}~\bibnamefont {Krasilnikov}}, \emph {et~al.},\ }\bibfield  {title}
  {\bibinfo {title} {Generating low transverse emittance beams for linac based
  light sources at pitz},\ }\href@noop {} {\bibfield  {journal} {\bibinfo
  {journal} {Proceedings of IPAC10, Kyoto, Japan}\ } (\bibinfo {year}
  {2010})}\BibitemShut {NoStop}%
\bibitem [{\citenamefont {Krasilnikov}\ \emph {et~al.}(2005)\citenamefont
  {Krasilnikov}, \citenamefont {Bahr}, \citenamefont {Grabosch}, \citenamefont
  {Han}, \citenamefont {Miltchev}, \citenamefont {Oppelt}, \citenamefont
  {Petrosyan}, \citenamefont {Staykov}, \citenamefont {Stephan},\ and\
  \citenamefont {Hartrott}}]{krasilnikov2005beam}%
  \BibitemOpen
  \bibfield  {author} {\bibinfo {author} {\bibfnamefont {M.}~\bibnamefont
  {Krasilnikov}}, \bibinfo {author} {\bibfnamefont {J.}~\bibnamefont {Bahr}},
  \bibinfo {author} {\bibfnamefont {H.-J.}\ \bibnamefont {Grabosch}}, \bibinfo
  {author} {\bibfnamefont {J.}~\bibnamefont {Han}}, \bibinfo {author}
  {\bibfnamefont {V.}~\bibnamefont {Miltchev}}, \bibinfo {author}
  {\bibfnamefont {A.}~\bibnamefont {Oppelt}}, \bibinfo {author} {\bibfnamefont
  {B.}~\bibnamefont {Petrosyan}}, \bibinfo {author} {\bibfnamefont
  {L.}~\bibnamefont {Staykov}}, \bibinfo {author} {\bibfnamefont
  {F.}~\bibnamefont {Stephan}},\ and\ \bibinfo {author} {\bibfnamefont
  {M.}~\bibnamefont {Hartrott}},\ }\bibfield  {title} {\bibinfo {title}
  {Beam-based procedures for rf guns},\ }in\ \href@noop {} {\emph {\bibinfo
  {booktitle} {Proceedings of the 2005 Particle Accelerator Conference}}}\
  (\bibinfo {organization} {IEEE},\ \bibinfo {year} {2005})\ pp.\ \bibinfo
  {pages} {967--969}\BibitemShut {NoStop}%
\bibitem [{\citenamefont {Floettmann}()}]{astracode}%
  \BibitemOpen
  \bibfield  {author} {\bibinfo {author} {\bibfnamefont {K.}~\bibnamefont
  {Floettmann}},\ }\href@noop {} {\bibinfo {title} {Astra: A space charge
  tracking algorithm}},\ \bibinfo {note}
  {\url{https://www.desy.de/~mpyflo/}}\BibitemShut {NoStop}%
\bibitem [{\citenamefont {Chen}\ \emph {et~al.}(2020)\citenamefont {Chen},
  \citenamefont {Zagorodnov},\ and\ \citenamefont {Dohlus}}]{chen2020beam}%
  \BibitemOpen
  \bibfield  {author} {\bibinfo {author} {\bibfnamefont {Y.}~\bibnamefont
  {Chen}}, \bibinfo {author} {\bibfnamefont {I.}~\bibnamefont {Zagorodnov}},\
  and\ \bibinfo {author} {\bibfnamefont {M.}~\bibnamefont {Dohlus}},\
  }\bibfield  {title} {\bibinfo {title} {Beam dynamics of realistic bunches at
  the injector section of the european x-ray free-electron laser},\ }\href@noop
  {} {\bibfield  {journal} {\bibinfo  {journal} {Physical Review Accelerators
  and Beams}\ }\textbf {\bibinfo {volume} {23}},\ \bibinfo {pages} {044201}
  (\bibinfo {year} {2020})}\BibitemShut {NoStop}%
\bibitem [{\citenamefont {Scholz}(2020)}]{Xfel250pC2020}%
  \BibitemOpen
  \bibfield  {author} {\bibinfo {author} {\bibfnamefont {M.}~\bibnamefont
  {Scholz}},\ }\bibfield  {title} {\bibinfo {title} {Beam optics and emittance
  measurement evaluations 2019 and 2020}} (\bibinfo {year} {2020}),\ \bibinfo
  {note} {{XFEL} Beam dynamics meeting}\BibitemShut {NoStop}%
\bibitem [{\citenamefont {Chen}\ \emph {et~al.}(2021)\citenamefont {Chen},
  \citenamefont {Brinker}, \citenamefont {Decking}, \citenamefont {Scholz},\
  and\ \citenamefont {Winkelmann}}]{chen2021perspectives}%
  \BibitemOpen
  \bibfield  {author} {\bibinfo {author} {\bibfnamefont {Y.}~\bibnamefont
  {Chen}}, \bibinfo {author} {\bibfnamefont {F.}~\bibnamefont {Brinker}},
  \bibinfo {author} {\bibfnamefont {W.}~\bibnamefont {Decking}}, \bibinfo
  {author} {\bibfnamefont {M.}~\bibnamefont {Scholz}},\ and\ \bibinfo {author}
  {\bibfnamefont {L.}~\bibnamefont {Winkelmann}},\ }\bibfield  {title}
  {\bibinfo {title} {Perspectives towards sub-{\aa}ngstr{\"o}m working regime
  of the european x-ray free-electron laser with low-emittance electron
  beams},\ }\href@noop {} {\bibfield  {journal} {\bibinfo  {journal} {Applied
  Sciences}\ }\textbf {\bibinfo {volume} {11}},\ \bibinfo {pages} {10768}
  (\bibinfo {year} {2021})}\BibitemShut {NoStop}%
\bibitem [{\citenamefont {Prat}\ \emph
  {et~al.}(2014{\natexlab{b}})\citenamefont {Prat}, \citenamefont {Bettoni},
  \citenamefont {Braun}, \citenamefont {Divall}, \citenamefont {Ganter},
  \citenamefont {Schietinger}, \citenamefont {Trisorio},\ and\ \citenamefont
  {Vicario}}]{prat2014thermal}%
  \BibitemOpen
  \bibfield  {author} {\bibinfo {author} {\bibfnamefont {E.}~\bibnamefont
  {Prat}}, \bibinfo {author} {\bibfnamefont {S.}~\bibnamefont {Bettoni}},
  \bibinfo {author} {\bibfnamefont {H.}~\bibnamefont {Braun}}, \bibinfo
  {author} {\bibfnamefont {M.}~\bibnamefont {Divall}}, \bibinfo {author}
  {\bibfnamefont {R.}~\bibnamefont {Ganter}}, \bibinfo {author} {\bibfnamefont
  {T.}~\bibnamefont {Schietinger}}, \bibinfo {author} {\bibfnamefont
  {A.}~\bibnamefont {Trisorio}},\ and\ \bibinfo {author} {\bibfnamefont
  {C.}~\bibnamefont {Vicario}},\ }\bibfield  {title} {\bibinfo {title} {Thermal
  emittance measurements at the swissfel injector test facility},\ }in\
  \href@noop {} {\emph {\bibinfo {booktitle} {These Proceedings: Proc. 36th
  Int. Free-Electron Laser Conf., Basel, Switzerland}}}\ (\bibinfo {year}
  {2014})\BibitemShut {NoStop}%
\bibitem [{\citenamefont {Qian}\ \emph {et~al.}(2021)\citenamefont {Qian},
  \citenamefont {Mohanty}, \citenamefont {Aboulbanine}, \citenamefont
  {Adhikari}, \citenamefont {Aftab}, \citenamefont {Boonpornprasert},
  \citenamefont {Good}, \citenamefont {Gross}, \citenamefont {Hoffmann},
  \citenamefont {Koschitzki}, \citenamefont {Krasilnikov}, \citenamefont
  {Lueangaramwong}, \citenamefont {Lishilin}, \citenamefont {Oppelt},
  \citenamefont {Niemczyk}, \citenamefont {Stephan}, \citenamefont {Shu},
  \citenamefont {Weilbach}, \citenamefont {Loisch}, \citenamefont {Chen},
  \citenamefont {Monaco}, \citenamefont {Sertore},\ and\ \citenamefont
  {Guerini}}]{Qian:471884}%
  \BibitemOpen
  \bibfield  {author} {\bibinfo {author} {\bibfnamefont {H.}~\bibnamefont
  {Qian}}, \bibinfo {author} {\bibfnamefont {S.}~\bibnamefont {Mohanty}},
  \bibinfo {author} {\bibfnamefont {Z.}~\bibnamefont {Aboulbanine}}, \bibinfo
  {author} {\bibfnamefont {G.~D.}\ \bibnamefont {Adhikari}}, \bibinfo {author}
  {\bibfnamefont {N.}~\bibnamefont {Aftab}}, \bibinfo {author} {\bibfnamefont
  {P.}~\bibnamefont {Boonpornprasert}}, \bibinfo {author} {\bibfnamefont
  {J.~D.}\ \bibnamefont {Good}}, \bibinfo {author} {\bibfnamefont
  {M.}~\bibnamefont {Gross}}, \bibinfo {author} {\bibfnamefont
  {A.}~\bibnamefont {Hoffmann}}, \bibinfo {author} {\bibfnamefont
  {C.}~\bibnamefont {Koschitzki}}, \bibinfo {author} {\bibfnamefont
  {M.}~\bibnamefont {Krasilnikov}}, \bibinfo {author} {\bibfnamefont
  {A.}~\bibnamefont {Lueangaramwong}}, \bibinfo {author} {\bibfnamefont
  {O.}~\bibnamefont {Lishilin}}, \bibinfo {author} {\bibfnamefont
  {A.}~\bibnamefont {Oppelt}}, \bibinfo {author} {\bibfnamefont
  {R.}~\bibnamefont {Niemczyk}}, \bibinfo {author} {\bibfnamefont
  {F.}~\bibnamefont {Stephan}}, \bibinfo {author} {\bibfnamefont
  {G.}~\bibnamefont {Shu}}, \bibinfo {author} {\bibfnamefont {T.}~\bibnamefont
  {Weilbach}}, \bibinfo {author} {\bibfnamefont {G.}~\bibnamefont {Loisch}},
  \bibinfo {author} {\bibfnamefont {Y.~L.}\ \bibnamefont {Chen}}, \bibinfo
  {author} {\bibfnamefont {L.}~\bibnamefont {Monaco}}, \bibinfo {author}
  {\bibfnamefont {D.}~\bibnamefont {Sertore}},\ and\ \bibinfo {author}
  {\bibfnamefont {G.}~\bibnamefont {Guerini}},\ }\bibfield  {title} {\bibinfo
  {title} {{F}irst {G}reen {C}athode {T}est in the {H}igh {G}radient {RF} {G}un
  at {PITZ}}\ }(\bibinfo {organization} {Photocathode Physics for
  Photoinjectors Workshop, California (USA), 10 Nov 2021 - 12 Nov 2021},\
  \bibinfo {year} {2021})\BibitemShut {NoStop}%
\end{thebibliography}%

\end{document}